\documentclass[10pt,aps,prd,notitlepage,superscriptaddress,nofootinbib,amsmath,amssymb,showpacs,twocolumn]{revtex4-1} %twocolumn
\usepackage[latin9]{inputenc}
\usepackage{aas_macros}

\pdfoutput=1

\usepackage{color,marvosym}
\usepackage{graphicx} 
\usepackage[normalem]{ulem}

\definecolor{darkred}{rgb}{0.5,0,0}
\definecolor{darkgreen}{rgb}{0,0.5,0}
\definecolor{darkblue}{rgb}{0,0,0.5}
\definecolor{orangeish}{rgb}{1,0.5,0}

\usepackage{hyperref}
\hypersetup{colorlinks,
linkcolor=darkblue,
filecolor=darkgreen,
urlcolor=darkblue,
citecolor=darkblue
}

\newcommand{\fett}[1]{\boldsymbol{#1}}
\newcommand{\dd}{{\rm{d}}}
\newcommand{\ii}{{\rm{i}}}
\newcommand{\e}{{\rm e}}
\newcommand{\be}{\begin{equation}}
\newcommand{\ee}{\end{equation}}
\renewcommand{\v}[1]{\boldsymbol{#1}}
\newcommand{\vx}{\v{x}}
\newcommand{\vv}{\v{v}}
\newcommand{\vp}{\v{p}}
\newcommand{\vq}{\v{q}}

\newcommand{\vnabla}{\v{\nabla}}
\newcommand{\vnablax}{\v{\nabla}_{\!x\,}}

\newcommand{\new}[1]{#1}

\makeatother

\begin{document}

\title{Semiclassical path to cosmic large-scale structure}

\author{Cora Uhlemann}
\email{c.uhlemann@damtp.cam.ac.uk}
\affiliation{Centre for Theoretical Cosmology, DAMTP, University of Cambridge, CB3 0WA Cambridge, United Kingdom}
\affiliation{Fitzwilliam College, University of Cambridge, CB3 0DG Cambridge, United Kingdom}

\author{Cornelius Rampf}
\email{cornelius.rampf@oca.eu} 
\altaffiliation{Marie Sk\l odowska--Curie Fellow}
\affiliation{Laboratoire Lagrange, Universit\'e C\^ote d'Azur, Observatoire de la C\^ote d'Azur, CNRS, Blvd de l'Observatoire, CS 34229, 06304 Nice, France}

\author{Mateja Gosenca}
\email{mateja.gosenca@auckland.ac.nz}
\affiliation{Department of Physics, University of Auckland, Private Bag 92019, Auckland, New Zealand}
\affiliation{Astronomy Centre, School of Mathematical and Physical Sciences,
University of Sussex, Brighton BN1 9QH, United Kingdom}

\author{Oliver Hahn}
\email{oliver.hahn@oca.eu}
\affiliation{Laboratoire Lagrange, Universit\'e C\^ote d'Azur, Observatoire de la C\^ote d'Azur, CNRS, Blvd de l'Observatoire, CS 34229, 06304 Nice, France}

\date{\today}

\begin{abstract}

We chart a path toward solving for the nonlinear gravitational dynamics of cold dark matter by relying on a semiclassical description using the propagator. The evolution of the propagator is given by a Schr\"odinger equation, where the small parameter $\hbar$ acts as a softening scale that regulates singularities at shell-crossing. 
The leading-order propagator, called free propagator, is the semiclassical equivalent of the Zel'dovich approximation (ZA), that describes inertial particle motion along straight trajectories. At next-to-leading order, we solve for the propagator perturbatively and obtain, in the classical limit the displacement field from second-order Lagrangian perturbation theory (LPT). The associated velocity naturally includes an additional term that would be considered as third order in LPT. We show that this term is actually needed to preserve the underlying Hamiltonian structure, and ignoring it could lead to the spurious excitation of vorticity in certain implementations of second-order LPT. 
\new{We show that for sufficiently small $\hbar$  the corresponding
propagator solutions closely resemble LPT, with the additions that spurious vorticity is avoided and the dynamics at shell-crossing is regularised. Our analytical results possess a symplectic structure that allows us to advance numerical schemes for the large-scale structure.}
For times shortly after shell-crossing, we explore the generation of vorticity, which in our method does not involve any explicit multi-stream averaging, but instead arises naturally as a conserved topological charge.

\end{abstract}

%\pacs{98.80.Es, 04.50.Kd, 95.36.+x}

\maketitle

%%%%%%%%%%%%%%%%%%%%%%%%%%%%%%%%%%%%%%%%%%%%%%%%%%%%
%%%%%%%%%%%%%%%%%%%%%%%%%%%%%%%%%%%%%%%%%%%%%%%%%%%%

\section{Introduction}
Analytical methods for the evolution of cold dark matter (CDM), relevant for investigating the large-scale structure of our Universe,
are valuable for gaining theoretical understanding and efficiently parametrising the cosmology-dependence of observables.
Furthermore, analytical methods can also assist in improving numerical computations, for example by setting up initial conditions for cosmological $N$-body 
simulations \cite{Hahn:2011uy,Crocce:2006ve}, or as input for multi-time-stepping algorithms that are used to generate fast mock catalogues \cite{Tassev13COLA,Feng:2016yqz}.

At sufficiently early times, the distribution for standard cold dark matter
is characterised by the (perfect) fluid variables, a density
and a single-valued velocity field. 
\new{These fields are the dynamical quantities in the Eulerian description. 
In the corresponding perturbative framework, Eulerian perturbation theory \cite{BernardeauReview},
one assumes a perturbative smallness of the density and velocity from their background values.
As a consequence,} Eulerian perturbation theory struggles to capture large densities that naturally arise from gravitational dynamics.
Moreover, due to the collisionless nature of dark matter,  
the infall induced by the gravitational potential leads to 
the crossing of fluid trajectories, which
is often called `shell-crossing'. 
The instance of shell-crossing leads to the formation of caustics with a (formally) infinite density.
Such non-analytic behaviour prevents a smooth continuation into the regime beyond shell-crossing, 
in which regions of multiple fluid streams with distinct velocities arise.

%At sufficiently early times, the distribution for standard cold dark matter
%is characterised by the (perfect) fluid variables, a density
%and a single-valued velocity field. 
%In the Eulerian description, one can perturbatively solve for the density and velocity 
%and provide recursive relations for higher-order solutions depending on 
%linear fields. However, Eulerian perturbation theory \cite{BernardeauReview} struggles to capture large densities that naturally arise from gravitational dynamics.
%Moreover, due to the collisionless nature of dark matter,  
%the infall induced by the gravitational potential leads to 
%the crossing of fluid trajectories, which
%is often called `shell-crossing'. 
%The instance of shell-crossing leads to the formation of caustics with a (formally) infinite density.
%Such non-analytic behaviour prevents a smooth continuation into the regime beyond shell-crossing, 
%in which regions of multiple fluid streams with distinct velocities arise.
%

Some problems associated with shell-crossing are alleviated in a phase-space description, where the
dark matter evolution corresponds to describing the embedding of 
an initially flat, 3-dimensional thin sheet 
into  6-dimensional phase-space, the `Lagrangian submanifold'. 
In this case, the singular behaviour of the density at shell-crossing does not pose a problem per se, 
because it is merely induced by projecting the folded phase-space sheet into 3-dimensional position space. 
A convenient way of describing the embedding of the sheet in phase-space is to use Lagrangian coordinates \cite{Villone2017,Zeldovich:1969sb,Buchert:1989xx,Bouchet:1992uh,Buchert:1993xz,Rampf:2012xa}. 
The central quantity is the Lagrangian displacement field that encodes how fluid elements are displaced as a function of time and initial (Lagrangian) position.
The corresponding perturbative framework is usually called Lagrangian perturbation theory (LPT), and, although a challenge, allows to
investigate the instance of shell-crossing  by analytic means \cite{Zheligovsky:2013eca,Rampf:2015mza,Taruya17,Rampf17,Rampf:2017tne,Saga18}. 
However, to update the gravitational potential that is responsible for displacing the fluid elements, one still requires, effectively, the Eulerian density.
Since an irregular density carries over to irregularities in the \new{tidal or force field~\cite{Pietroni:2018ebj}, a theory that regularises the dynamics at shell-crossing and provides a smooth transition into the multi-stream regime
is highly desirable.}

In this paper, we provide a novel method to evolve CDM by relying on semiclassical dynamics using 
the propagator -- which we motivate in more detail in Sec.\,\ref{sec:propagator}. 
\new{Our primary motivation is to formulate an analytical framework that allows to go beyond shell-crossing while admitting perturbative solutions for the mildly nonlinear dynamics. To realise this goal, we adopt a formalism that respects key dynamical invariants, which additionally offers interesting perspectives for advancing numerical schemes for gravitational dynamics.}

In essence, we introduce a suitable perturbative framework for the propagator and show that, at the leading order,
our approach amounts to solving a `free-particle Schr\"odinger equation', 
with a solution that is closely related to the classical ZA dynamics.
At the leading order, our approach constitutes a heuristic derivation for a free-particle Schr\"odinger equation, which has been introduced in Refs.\,\cite{ColesSpencer03,ShortColes06wave}.
At higher orders, our approach includes gravitational effects beyond the ZA,
which are encoded in an effective potential. The resulting evolution equation for the propagator
is then a Schr\"odinger equation that includes the said effective potential. 
Our formalism naturally respects the underlying Hamiltonian structure
and allows us to determine the regularised fluid variables in the classical limit.
Additionally, our formalism does not suffer from singularities at shell-crossing, 
thereby allowing it to enter the multi-stream regime smoothly.

The outline of the paper is as follows. 
In Sec.\,\ref{sec:propagator}, we begin by motivating our semiclassical description, which at the leading order returns a free propagator that reduces to the ZA in the classical limit.
We then generalise the related Schr\"odinger equation for the propagator to include an effective potential.
In Sec.\,\ref{sec:fluid}, we relate this effective potential to the cosmological fluid equations and describe standard perturbative solutions. 
In Sec.\,\ref{sec:LPT}, we briefly review LPT since it will be helpful to analyse the results of the propagator method. Specifically, we 
utilise the so-called Cauchy invariants, that in our case encode the conservation of vanishing vorticity, as a diagnostic tool to investigate the Hamiltonian structure of the propagator method. 
In Sec.\,\ref{sec:Observables}, we show how the classical limits for the propagator relate to the classical fluid variables. As a demonstrative example, we perform those limits on the free propagator, from which we reproduce the ZA.
Furthermore, we discuss how vorticity is generated by shell-crossing in the semiclassical picture.
In Sec.\,\ref{sec:PPT}, we go to the next-to-leading order in the propagator method and show that we recover exactly the second-order Lagrangian displacement. We show that the associated velocity 
receives a higher-order correction term compared to LPT, restoring compatibility with the Cauchy invariants thanks to the underlying symplectic structure of our method.
In Sec.\,\ref{sec:Examples}, we present concrete examples comparing numerical implementations of our propagator method to LPT, and show qualitative results for the density and vorticity after shell-crossing. 
We conclude in Sec.\,\ref{sec:conclusion}, where we also provide an outlook on potential applications of our method.

%%%%%%%%%%%%%%%%%%%%%%%%%%%%%%%%%%%%%%%%%%%%%%%%%%%%%%%%

\section{Motivating a propagator for gravitational interactions}
\label{sec:propagator}

Despite its simplicity, the classical Zel'dovich approximation (ZA)~\cite{Zeldovich:1969sb}  
reproduces the cosmic web remarkably well.
The ZA is based on a ballistic motion of fluid  
particles with prescribed velocity. 
While the ZA is the exact solution of the underlying fluid equations before shell-crossing in one spatial dimension, it receives corrections due to tidal gravitational fields in two and three dimensions. 
Furthermore, after the intersection of 
particle trajectories at shell-crossing, the ZA is doomed to fail, partially because
it does not take the time evolution of the gravitational potential into account, which leads to nonzero particle acceleration and secondary infall.

In the following, we develop another perspective for the ZA, and outline how this model 
can be translated into the language of transition probabilities  
-- with the aim to investigate novel avenues 
to surpass some of the shortcomings of the ZA. We shall make use of the dimensionless temporal coordinate $a = a(t)$, which is 
the cosmic scale factor with evolution governed by the usual Friedmann equations. 
For simplicity, we shall limit the analysis to a CDM dominated universe, the so-called Einstein--de Sitter (EdS) model. A generalisation to the $\Lambda$CDM model, 
where $\Lambda$ is the cosmological constant, is however straightforward.

\subsection{Free propagator from Zel'dovich approximation}
In a Cartesian coordinate system which comoves with the particles, the ZA amounts to particles moving along straight lines with 
constant velocity. For fluid particles that are initially ($a=0$) at position $\vq$, and at time $a$, at position $\vx$,
the classical action is
\be
\label{eq:S0}
S_0(\vx,\vq; a) = \frac{1}{2} (\vx-\vq) \cdot\frac{\vx-\vq}{a}\,, 
\ee
which, essentially, is the product of particle displacements and their velocities. 
Following standard methods inspired by Feynman \cite{FeynmanHibbs} 
the classical action yields a transition amplitude $K_0$ for a particle from being at time $0$ at position $\vq$ to being at time $a$ at position $\vx$, 
\be
\label{eq:K0sol}
K_0(\vx,\vq;a) = N
   \exp\left\{ \frac{\ii}{\hbar} S_0(\vx,\vq; a) \right\} \,,
\ee
where $N =  (2 \pi \ii \hbar a)^{-3/2}$ is a normalisation coefficient chosen such that the transition amplitude returns initially the Dirac delta,
$K_0(\vx, \vq; \text{{\small $a\!=\!0$}})=\delta_{\rm D}^{(3)}(\vx-\vq)$.

Now, as is well known in the context of quantum mechanics,  the transition amplitude $K_0$  (and any other transition amplitude too) propagates a wave function $\psi_0$ from some initial to some final state,
\be
\label{eq:PsiKernel}
  \psi_0(\vx;a) =  \int \! \dd^3 q \, K_0(\vx,\vq;a) \,  \psi_0^{\rm (ini)}(\vq) \,,
\ee 
where $\psi_0^{\rm (ini)}(\vq) \equiv \psi_0(\vq; \text{\small $a=0$})$.
It is then elementary to show that,  
the evolution of the transition amplitude $K_0$ is governed by a potential-free Schr\"odinger equation,
\begin{subequations}
\begin{align} \label{eq:freeSchroedi}
  \ii \hbar \partial_a K_0 &= -\frac{\hbar^2}{2} \fett{\nabla}_x^2 K_0 
\intertext{\new{for $a>0$} and, likewise,}
 \ii \hbar \partial_a \psi_0 &= \ - \frac{\hbar^2}{2} \fett{\nabla}_x^2 \psi_0\,,
\end{align}
\end{subequations}
which is a free Schr\"odinger equation for the ZA wave function. 
This provides a heuristic derivation for the `free particle approximation', which was introduced as an {\it ad-hoc} approximation of gravitational dynamics in \cite{ColesSpencer03}\footnote{It appears that the work of~\cite{ColesSpencer03}
was inspired by the Schr\"odinger-Poisson description of~\cite{Widrow1993}, however, as we explain in App.\,\ref{app:Schroedi}, the two approaches are formally not equivalent.}, 
and compared to the linearised fluid dynamics in \cite{ShortColes06wave}.

Even at the leading order, our approach goes beyond the so-called adhesion model \cite{Gurbatov89,Weinberg90}, 
which avoids shell-crossing through the introduction of an artificial viscosity, and has also been formulated in terms of propagators \cite{Jones99,Matarrese02,Chavanis11,Rigopoulos15}. 
In contrast to the adhesion approximation, our semiclassical propagator method possesses a
conserved current that \new{arises from the unitary evolution of the wave function and} allows us to propagate through shell-crossing into the multi-stream regime.

\begin{figure}
\includegraphics[width=0.92\columnwidth]{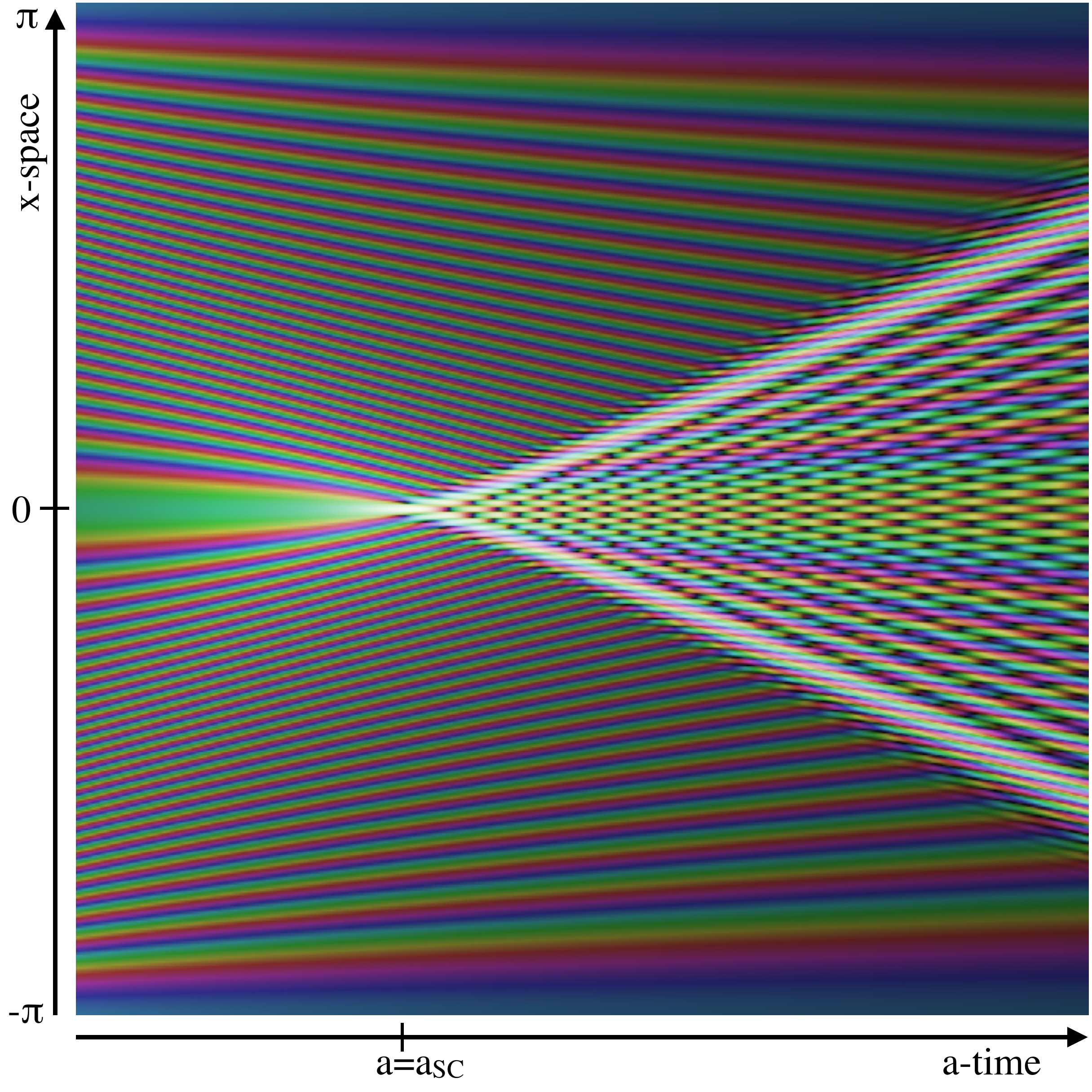}
\caption{Time evolution of the wave function $\psi_0(x;a)$ with 1D initial data $\psi_0^{\rm (ini)}(q) = \exp\{   (\ii/ \hbar ) \cos q\}$,
with $\hbar = 0.01$,
and evolved using a grid of $1024$ cells.
Using domain coloring, the figure shows both the amplitude of the wave function (corresponding to the square root of the density) in terms of the brightness as $0.5^{|\psi|}$, as well as the phase in terms of the color hue. Lines of constant color thus correspond to trajectories of constant phase. 
For times $a \geq a_{\rm sc}$, an interference pattern arises as a result of multi-streaming. 
} \label{fig:free-sine-wave1}
\end{figure}

The above equation is easily solved either by numerical or analytical means (see the following sections for details). As a demonstrative example, Fig.\,\ref{fig:free-sine-wave1}
shows the evolution of the wave function $\psi_0$ with the 1D initial data $\psi_0^{\rm (ini)}(q)\!=\!\exp\left( - \ii \phi_{\rm v}^{\rm (ini)}/\hbar\right)$, with $\phi_{\rm v}^{\rm (ini)}\!=-\cos q$.
The graph uses the domain coloring technique to assign a unique color to each point of the complex plane (cf.\ \cite{Wegert:2012}), so that the amplitude of $\psi_0$ is mapped to the brightness and the phase to the color hue for each point in the space-time plane $(x,a)$.

From Fig.\,\ref{fig:free-sine-wave1} it becomes evident that, from the time of shell-crossing $a_{\rm sc}$ onwards, interference patterns arise. 
These patterns can be understood from a combined perspective based on fluid dynamics and wave mechanics:
waves resemble fluid particles, and when individual fluid trajectories cross, the waves  
create interference patterns. The crossing of particle trajectories implies the transition from single to multi-stream regime.   
In cosmological fluid dynamics, this instance is often denoted with shell-crossing 
\new{with the appearance of infinite densities~\cite{Arnold1982,Hidding2014,Feldbrugge:2017ivf}.}
In the wave-function approach, by contrast, infinite densities are naturally 
regularised by a nonzero~$\hbar$ which acts as a softening scale. 
The particular semiclassical diffraction pattern emerging here from a classical caustic (called a `fold' or `cusp') is also known from wave optics, in particular in the framework of catastrophe optics~\cite{Berry1980}.

Furthermore, Fig.~\ref{fig:free-sine-wave1} displays the dispersion of the outward propagating caustics, emanating around $a = a_{\rm sc}$ and $x=0$.
This smearing effect is expected from the quantum dispersion relation, and leads to an uncertainty in the space location and time of shell-crossing for $\hbar >0$.

\begin{figure*}
\includegraphics[width=0.9\textwidth]{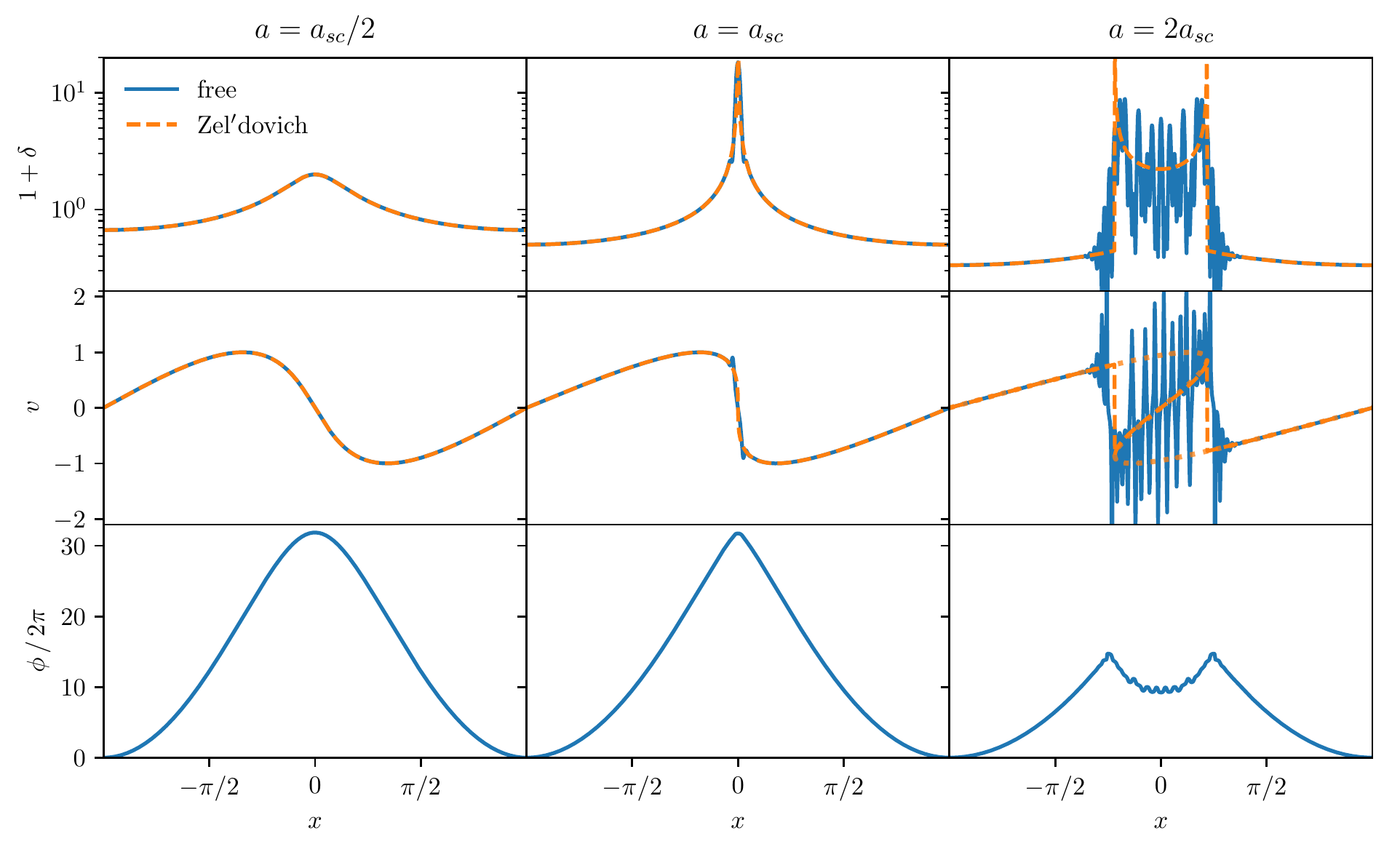}
\caption{Time-evolution of the density (upper row panels), velocity (middle row) as well as the phase of the wave function (lower row) for the plane-wave collapse, evolved using the free propagator (solid blue lines) with $\hbar=0.02$ on a grid of $1024$ cells, and the Zel'dovich approximation (dashed orange lines). Results are shown at three different times: well before shell-crossing at $a=a_{\rm sc}/2$, at shell-crossing $a=a_{\rm sc}$, and well after shell-crossing at $a=2 a_{\rm sc}$. For the Zel'dovich case, we show after shell-crossing both the mean velocity (dashed line), computed from the first moment of the distribution function, as well as the full phase-space curve (dotted line). These two would be identical up to the instant of shell-crossing.} 
\label{fig:free-sine-wave2}
\end{figure*}

In Fig.\,\ref{fig:free-sine-wave2} we show the density and velocity associated with the free wave function~\eqref{eq:PsiKernel}, and compare the results against the classical ZA.
We make use of the same 1D initial data and $\hbar$-values as before. From the time-evolution of the wave function, written as $\psi=\sqrt{1+\delta}\exp(-\ii\phi_{\rm v}/\hbar)$, we extract the normalised density $1+\delta$, the velocity $v= -\nabla \phi_{\rm v}$ and 
phase $\phi_{\rm v}$. Before shell-crossing ($a\!=\!a_{\rm sc}/2$), the wave approach agrees well with the ZA, while at the time of (classical) shell-crossing, when the density in the ZA becomes infinite, the wave-density remains finite thanks to the taming scale $\hbar$. After shell-crossing ($a\!=\!2 a_{\rm sc}$), the ZA leads to the known overshooting in the multi-stream regime---which is most clearly seen in the velocity plot---while the wave approach leads to interference patterns. \new{Since the semiclassical method relies on a finite phase-space resolution associated with the small parameter $\hbar$, the density and velocity associated with the wave-function should be interpreted in a coarse-grained sense, which washes out small-scale oscillations; see Sec.\,\ref{sec:Observables} for further details.}

\subsection{Propagators with an effective potential}
\label{sec:propeffpot}

Beyond the ZA, which is the exact solution of the one-dimensional collapse problem before shell-crossing,
particles will not move along straight trajectories but will 
be influenced by gravitational tidal effects. 
To transmit such tidal interactions, an effective (gravitational) potential should be included in the Hamiltonian operator.
Consequently, 
the \new{associated} propagator $K(\vx,\vq;a)$ and wave function $\psi(\vx;a)$ satisfy respectively 
the following Schr\"odinger equations \new{($a>0$)}
\begin{align}
\label{eq:Schroedi}
 \ii \hbar \partial_a K &= \hat H K  \,, \qquad \qquad  \ii \hbar \partial_a \psi  =\hat H \psi \,, 
\end{align}
with the Hamiltonian operator
\be
\label{eq:Hop}
  \hat H \equiv - \frac{\hbar^2}{2} \fett{\nabla}_x^2  + V_{\rm eff}(\vx;a) \,,
\ee
where $V_{\rm eff}$ can be considered as an external potential. An explicit expression for the effective potential is given in Eq.\,\eqref{eq:Veffdef}.
We remark that while this formulation indeed can capture both the effects from going beyond the ZA and beyond shell-crossing, in the present paper we focus on the dynamics before (and at) shell-crossing. \new{For a discussion of the relation of our semiclassical method to other quantum-inspired formalisms relying on a Schr\"odinger equation we refer to App.\,\ref{app:Schroedi}.}

\subsection{Strategy for solving the propagator equation} 

The aim of this paper is to investigate perturbative solutions for the Schr\"odinger equation~\eqref{eq:Schroedi}.
For this, we discuss the following aspects of our propagator formalism in the context of perturbative schemes.

\begin{enumerate}
  \item  As an external potential for the Hamiltonian operator $\hat H$ in Eq.\,\eqref{eq:Hop},
     we employ an effective potential that is determined by using standard perturbative techniques for the classical fluid equations, see Sec.\,\ref{sec:fluid}.    
  
  \item As a benchmark for our perturbative results, we will use LPT and the Cauchy invariants as diagnostic tool for detecting spurious effects in perturbative truncations, which we introduce in Sec.\,\ref{sec:LPT}.

\item  We show how quantities constructed from the wave function are related to  the classical fluid variables, by employing classical limits; see Sec.\,\ref{sec:Observables}.

\item Finally, in Sec.\,\ref{sec:PPT}, we provide a perturbative expansion for the propagator, solve the problem up to the next-to-leading order (NLO), and determine the associated fluid variables in the classical limit.

\end{enumerate}

%%%%%%%%%%%%%%%%%%%%%%%%%%%%%%%%%%%%%%%%%%%%%%%%%%%%%%%%
\section{The cosmological fluid equations and the effective potential}
\label{sec:fluid}

In this section, we relate the external potential in the Schr\"odinger equation~\eqref{eq:Schroedi} to an effective potential within the cosmological fluid equations. We show that, in the fluid description, this effective potential is a combination of the gravitational potential and a term due to the overall expansion of the Universe.

On sufficiently large scales and before shell-crossing, dark matter can be treated as a perfect fluid described in terms of a single-valued velocity and density. The  corresponding
equations are usually formulated in comoving coordinates $\vx = \v{r}/a$, where $\v{r}$ is the physical space coordinate and $a$ the cosmic scale factor. 
For convenience, we decompose the fluid density $\rho(\vx;a)$ into a background part $\bar \rho(a)$ and a density contrast $\delta$, which are related to each other via $\delta = (\rho - \bar \rho)/ \bar \rho$.
Furthermore, we make use of a peculiar velocity $\vv=\dd\vx/\dd a$ which is related to the total velocity via $\fett{U} =  H \v{r}  +  H a^2 \vv$, where $H$ is the Hubble parameter.
In the present work, we restrict our analysis to
a spatially flat universe solely filled with cold dark matter, the so-called Einstein--de Sitter (EdS) universe. A generalisation to a $\Lambda$CDM Universe (and beyond)
is however straightforward.

For the case of a potential velocity $\vv\equiv - \vnabla \phi_{\rm v} $, 
the fluid equations in an Eulerian coordinate system consist of the Bernoulli, continuity and Poisson equation, which are (see e.g.\ \cite{Rampf:2015mza}) 
\begin{subequations}
\label{eqs:fluid}
\begin{align}
  \partial_a  \phi_{\rm v}  - \frac{1}{2}  |\vnabla \phi_{\rm v}|^2 &= V_{\rm eff}\,,  \label{eq:Bernoulli}\\
  \partial_a \delta  -  \vnabla \cdot [ (1+\delta) \vnabla\phi_{\rm v} ] &= 0 \,, \label{eq:Conti} \\
 \nabla^2 \varphi_{\rm g} &= \frac \delta a\label{eq:Poisson}
    \,,
\end{align}
\end{subequations}
where \new{$\varphi_{\rm g}$ is the gravitational potential, and} we have defined the effective potential
\be
\label{eq:Veffdef}
  V_{\rm eff} \equiv \frac{3}{2a}  \left( \varphi_{\rm g} - \phi_{\rm v} \right) \,,
\ee
which is a combined potential that encapsulates the effects from the cosmological potential and the Hubble friction.

Formally linearising the fluid variables around its background
and evaluating the linearised fluid equations at arbitrary early times $a \to 0$, it is found that analytic solutions at $a=0$ exist only provided one makes use of the following boundary conditions
\be \label{slaving}
  \delta^{\rm (ini)} =0 \,, \qquad \qquad \varphi_{\rm g}^{\rm (ini)} = \phi_{\rm v}^{\rm (ini)} \,.
\ee
These boundary conditions select the growing-mode solutions, and are in accordance with our requirement of a potential velocity.

Equipped with these boundary conditions, it is straightforward to investigate the standard perturbation theory (SPT) for the fluid equations~\eqref{eqs:fluid}. The following power series {\it Ans\"atze} lead to simple recursion relations that consistently solve the fluid equations order by order,
\begin{align}
  \label{eq:Veffexpansion}
    \delta(\fett{x};a)  &= \sum_{n=1}^\infty \delta^{(n)}(\vx) \,a^{n} \,, \quad \phi_{\rm v}(\fett{x};a) = \sum_{n=1}^\infty \phi_{\rm v}^{(n)}(\vx) \,a^{n-1}   \nonumber \\
   V_{\rm eff}(\fett{x};a) &= \sum_{n=1}^\infty V_{\rm eff}^{(n)}(\vx) \,a^{n-2} \,.
\end{align}
All-order results for the density and velocity potential are well known (see e.g.\ \cite{BernardeauReview}), from which one can easily construct all-order perturbative results for the effective potential. Explicitly, at first order we have simply \mbox{$V_{\rm eff}^{(1)} = 0$,} whereas at second order we find
\be \label{eq:Veff}
V_{\rm eff}^{(2)} = \frac 3 7 \nabla^{-2} \left[ \left( \nabla^2\varphi_{{\rm g}}^{\rm (ini)} \right)^2  -  \left( \nabla_i \nabla_j \varphi_{\rm g}^{\rm (ini)} \right)^2 \right] \,,
\ee
with the inverse Laplacian $\nabla^{-2}$, and implied summation over dummy indices. For calculational details, including explicit all-order expressions \mbox{for $V_{\rm eff}^{(n)}$}, see App.\,\ref{app:Veff}.
Observe that $V_{\rm eff}^{(2)}$ is exactly zero for 1D initial conditions, in which case $\varphi_{\rm g}^{\rm (ini)}$ only depends on one spatial coordinate. In the Lagrangian-coordinates formulation, 
this is just the statement that the ZA is exact only in 1D; see the following section. 
Beyond 1D, $V_{\rm eff}^{(2)}$ is generally nonzero and constitutes the most important tidal correction to the leading order, and thus should be included for realistic modelling of the gravitational instability.

Before shell-crossing, one can relate the fluid picture to the Schr\"odinger equation, by using the Madelung polar form for the wave function $\psi=\sqrt{1+\delta}\exp(-\ii\phi_{\rm v}/\hbar)$ in Eq.\,\eqref{eq:Schroedi}. This way, one reproduces the fluid-type equations~\eqref{eqs:fluid}, with a Bernoulli equation~\eqref{eq:Bernoulli} that receives a new term $\sim \hbar^2$ \cite{Madelung27,ShortColes06schroedi}. Solving these fluid-type equations perturbatively with the {\it Ans\"atze}~\eqref{eq:Veffexpansion} reveals that $V_{\rm eff}$ agrees up to second order with the one obtained from the pure fluid approach, see App.~\ref{app:Veff}. Thus, up to second order in perturbation theory, which is our focus here, the effective potential can be identically obtained from either of the two approaches.

Equipped with an expression for the effective potential appearing in the Schr\"odinger equation~\eqref{eq:Schroedi},
we have the ingredients to perturbatively compute solutions for the propagator and hence the wave function. Before proceeding with 
the perturbative treatment of the propagator in Sec.\,\ref{sec:PPT}, let us provide a motivation 
based on Lagrangian perturbation theory in Sec.\,\ref{sec:LPT} and discuss how our semiclassical formalism can be used to infer classical observables in Sec.\,\ref{sec:Observables}.

\section{Lagrangian perturbation theory and Cauchy invariants}
\label{sec:LPT}

Perturbative solutions to the fluid equations can also be obtained in Lagrangian coordinates, and these solutions are determined within the framework of Lagrangian perturbation theory (LPT). 
Utilising a Lagrangian-coordinates approach has several advantages over the Eulerian approach.
From the theory side, 
there exists a mathematical proof of time-analyticity only for the Lagrangian-coordinates approach. LPT is therefore a convergent perturbation theory for the fluid equations and can solve them in the single-stream regime to arbitrary high accuracy \cite{Zheligovsky:2013eca,Rampf:2015mza}.

From the numerical side, $N$-body simulations of cosmic structure formation naturally initialise and evolve particles by employing coordinates that comove with the fluid.
In a combined effort to close the gap between theory and numerics, a Lagrangian-coordinates approach appears thus to be most suitable.

We denote by $\fett{q}$ the Lagrangian coordinates, and 
partial derivatives with respect to the component $q_i$ on a given function $f$ as $f_{,i}$ (or by $\nabla_i^{\rm L} f$). 
As before, summation over repeated indices is implied.
Let 
\be
\label{eq:defDisplacement}
  \fett{q} \mapsto \fett{x}(\fett{q};a) = \fett{q} + \fett{\xi}(\fett{q};a)
\ee
be the Lagrangian map from initial ($a\!=\!0$) position $\fett{q}$ to final Eulerian position $\fett{x}$ at time $a$, where $\fett{\xi}(\fett{q};a)$ is the Lagrangian displacement field. The map is defined in such a way that its Lagrangian (convective) $a$-time derivative $\partial_a^{\rm L}$, henceforth denoted by an overdot,
returns the Lagrangian representation of the velocity, i.e., 
\be 
\label{eq:defVelLag}
 \fett{v}(\fett{x}(\fett{q};a);a) = \partial_a^{\rm L} {\fett{x}} (\fett{q};a)  \equiv \dot{\fett{x}} (\fett{q};a) \,.
\ee
At initial time, the velocity is $\fett{v}^{\rm (ini)}(\fett{q}) = \fett{v}(\fett{x}(\fett{q};0);0)$ which agrees with the initial Eulerian velocity. 
Until the first shell-crossing, mass conservation $\bar \rho [1+\delta(\vx)]\dd^3x=\bar \rho\, \dd^3 q$ can be exactly integrated to give
\be
\label{eq:masscons}
  \delta = 1/ J -1 \,,
\ee
where $J=\det x_{i,j}$ is the determinant of the Jacobian matrix $x_{i,j}$.
Momentum conservation for the Lagrangian map can be conveniently written as
\begin{subequations} \label{eq:Lag}
\be \label{evoLag}
  \ddot{\fett{x}}(\fett{q};a) + \frac{3}{2a} \dot{\fett{x}}(\fett{q};a) = - \frac{3}{2a} \vnablax \varphi_{\rm g}(\fett{x}(\fett{q};a);a) \,,
\ee
which, after taking an Eulerian divergence and using the above definitions, turns into a scalar equation for the Lagrangian displacement field (see e.g.\,\cite{Rampf17}).

The considered flow is potential in Eulerian coordinates, which implies that the Eulerian curl of the Eulerian velocity vanishes, i.e., $\vnablax \!\!\times \vv\!=\!\fett{0}$.
By employing the Lagrangian map, the statement of zero vorticity translates in Lagrangian coordinates to the so-called Cauchy invariants, 
with components ($i=1,2,3$) \cite{Rampf:2012up,Zheligovsky:2013eca}
\be 
\label{eq:cauchy}
  \varepsilon_{ijk} \,  x_{l,j} \,\dot x_{l,k} = 0 \,,
\ee
\end{subequations}
where $\varepsilon_{ijk}$ is the Levi-Civita symbol. 
See~\cite{Matsubara:2015ipa} for a simple derivation of the Cauchy invariants which however holds only for zero vorticity; see~\cite{Rampf:2016wom} where initial vorticity is included. We note that in 1D, the Cauchy invariants are trivially satisfied, since in 1D there is no vorticity.

Equations~\eqref{eq:Lag} constitute a closed set of Lagrangian evolution equations, with all the dynamical information being encoded in the displacement field $\fett{\xi}$ (from which one can recover the density and velocity). 
To obtain perturbative solutions, one then writes for the displacement components
\be
   \xi_i(\fett{q};a) = \sum_{n=1}^\infty \xi_i^{(n)}(\fett{q})\, a^n \,,
\ee 
where the time-Taylor coefficients are given by simple all-order recursion relations \cite{Zheligovsky:2013eca}. For example, the first coefficients are
\begin{subequations}
 \label{eq:2LPTclass}
\begin{align}
\label{eq:ZA}
  \xi_i^{(1)} &=  - \nabla_i^{\rm L} \varphi_{\rm g}^{\rm (ini)}(\vq)  \hspace{0.04cm}\equiv \xi_i^{\rm ZA}(\vq;a) /a \,, \\
  \label{eq:2LPT}
  \xi_i^{(2)} &= - \frac 3 7 \nabla_{\rm L}^{-2} \nabla_i^{\rm L} \mu_2 \equiv \xi_i^{\rm 2LPT}(\vq;a) /a^2 \,, 
\end{align}
\end{subequations}
with the kernel
\be
\label{eq:mu2}
  \mu_2(\fett{q}) = \frac 1 2 \left[ \left( \nabla^2_{\rm L} \varphi_{\rm g}^{\rm (ini)} \right)^2 - \left(\nabla_i^{\rm L} \nabla_j^{\rm L} \varphi_{\rm g}^{\rm (ini)} \right)^2 \right] \,.
\ee
Observe the structural similarity of $\mu_2$ and $\nabla^2 V_{\rm eff}^{(2)}$  (equation~\eqref{eq:Veff}). At second order, the only difference between these two expressions is their Lagrangian vs.\ Eulerian coordinate dependences and derivatives.

The expressions $\xi_i^{\rm ZA}(\vq;a)$ and $\xi_i^{\rm 2LPT}(\vq;a)$ are the well-known Zel'dovich \cite{Zeldovich:1969sb,Buchert:1993xz} and 2LPT \cite{Buchert94,Bouchet:1994xp} approximations, respectively. The ZA, which is exact in one dimension, states that
fluid elements move along straight trajectories with a velocity set by the gradient of the initial gravitational potential $\varphi_{\rm g}^{\rm (ini)}$.
The 2LPT extension includes gravitational tidal effects, thereby taking into account the fact that gravity is non-local (beyond 1D).

Observe that both the Zel'dovich and 2LPT solutions give a displacement that is a gradient field and hence possesses a scalar potential. This property ceases to be true beyond second order, where one needs to include transverse modes, which is precisely the stage where one is forced to evaluate the Cauchy invariants~\eqref{eq:cauchy}.

\subsection{Spurious vorticity in perturbation theory}
\label{sec:fakevorticity}

While the Cauchy invariants are non-perturbative expressions, in LPT they translate into perturbative relations. 
At fixed order $n$ in LPT, we define the corresponding ``truncated'' Cauchy invariants by 
\be
\label{eq:cauchyLPT}
{\cal C}_i^{[n]} := \varepsilon_{ijk} \left(\delta_{lj}+\xi_{l,j}^{[n]}\right)\dot \xi_{l,k}^{[n]}  \,, 
\ee
where $\xi_{l}^{[n]} (\fett{q};a) := \sum_{i=1}^n \xi_{l}^{(i)}(\fett{q})\, a^i$ and similarly for $\dot \xi_{l}^{[n]}$. Even for vanishing vorticity, Eqs.\,\eqref{eq:cauchyLPT} are generically nonzero, as a consequence of the said truncation. Indeed, only at first order it is easily seen that ${\cal C}_i^{[1]}$ is still zero, but at the next truncated order we have 
\begin{align}
\label{eq:cauchy2LPT}
{\cal C}_i^{[2]}&= \varepsilon_{ijk} \left(\delta_{lj}+\xi^{\rm ZA}_{l,j}+\xi^{\rm 2LPT}_{l,j}\right) \left[\dot\xi^{\rm ZA}_{l,k}+\dot\xi^{\rm 2LPT}_{l,k}\right] \nonumber \\
&= \frac{a^2}{2} \varepsilon_{ijk} \varphi_{{\rm g},lj} V^{(2)}_{{\rm eff},lk} +\mathcal O(a^3)  \,,
\end{align}
where for later convenience we have expressed the Lagrangian kernel $\mu_2(\fett{q})$ in terms of $V_{\rm eff}^{(2)}(\fett{q})$.
This shows that the truncated Cauchy invariants are evidently nonzero at $\mathcal O(a^2)$,
and hence consistent with the Cauchy invariants in a perturbative sense.
However, those higher-order terms can excite unwanted spurious vorticity effects that are inconvenient when setting up initial conditions for $N$-body simulations.
Such unwanted spurious effects could be artificially amplified when a 2LPT-scheme with insufficient time-stepping is employed to generate fast mock simulations. Furthermore, from theoretical grounds it is expected that such spurious effects could be 
enhanced at late times. Indeed, the convergence radius of LPT, which should set the maximal step-size for such algorithms, depends on the inverse of the norm of the velocity gradients \cite{Zheligovsky:2013eca,Rampf:2015mza}. As a consequence, at late times, when velocity gradients become large, the convergence radius will naturally shrink. If a fixed time step is used in these algorithms that is larger than the radius of convergence, then the error in the velocity will grow as $\sim a^2$. \new{In the central panel of Fig.\,\ref{fig:vorticity-spurious} we provide numerical evidence of spurious vorticity generation by using 2LPT.}

In the subsequent Sec.\,\ref{sec:Observables}, we employ classical limits to relate the propagators and wave functions introduced in Sec.\,\ref{sec:propagator} to the Lagrangian fluid variables
 discussed here. As we will demonstrate explicitly in Sec.\,\ref{sec:PPT}, our  
propagator formalism avoids leading-order spurious vorticity effects. This is a consequence of the underlying Hamiltonian structure of the Schr\"odinger equation~\eqref{eq:Schroedi}, which ensures the conservation of certain integral invariants associated with vorticity, which we review next.

\subsection{Invariants associated with vorticity}
\label{sec:invariants}

Even for an initially potential flow, in two or higher dimensions, an effective vorticity will be generated beyond shell-crossing.
In classical collisionless dynamics, this effective vorticity arises through an averaging of the density weighted multi-stream velocities. 
The fact that vorticity can only arise due to multi-stream averaging 
is owed to the Helmholtz theorem for the conservation of vorticity flux 
(which can be linked to the more global Poincar\'e invariant). 
This theorem is closely related to the circulation theorem which is usually attributed to Lord Kelvin. 
Using the Cauchy invariants and Stoke's theorem, we obtain the (combined) Kelvin-Helmholtz theorem that states that~\cite{Frisch2017}
\be
\label{eq:Kelvin}
 \!\Gamma \equiv\oint_{C(a)} \!\v{v}\cdot \v{\dd x} =  \int_{S^{\rm (ini)}}  
\! (\fett{\nabla}^{\rm L} \times \fett{v}^{\rm (ini)}) \cdot \v{\dd S}^{\rm (ini)}
\ee
is an integral invariant, which is restricted to closed integral curves $C(a)$
in configuration space that follow the \new{inviscid} flow. Those curves describe the boundary of a surface, which initially is $S^{\rm (ini)}$ and has a corresponding oriented surface element $\v{\dd S}^{\rm (ini)}$.
The RHS of~\eqref{eq:Kelvin} is evaluated 
at initial time, which highlights that the integral invariant is 
a constant of motion. In the present case where there is no initial vorticity, $\Gamma =0$. 

The Cauchy invariants are essentially a local form of this integral invariant. The conservation of 
those invariants associated with vorticity, especially beyond shell-crossing, can also be understood from a fluid perspective.
For \new{an inviscid} fluid  that is at some earlier time in the single-stream regime and irrotational, the evolution equation for the displacement field is sourced by a gradient of the gravitational potential, which thus implies that there is no source of vorticity (see Eq.\,\eqref{evoLag}).
Now, after shell-crossing, when transiting from a single-fluid to a multi-fluid description,  
the evolution of each fluid stream is still governed by the same fluid-like evolution equation, with the only addition that the individual streams are now coupled gravitationally to the other streams through a common gravitational potential. However, crucially, the evolution equation for a given fluid stream is still sourced by a gradient of a gravitational potential. Thus, even in the multi-fluid regime, 
there is no source of vorticity, and 
each fluid stream remains potential at all times. However, as outlined above, there is generally an effective vorticity that results from averaging over the multiple streams.

%%%%%%%%%%%%%%%%%%%%%%%%%%%%%%%%%%%%%%%%%%%%%%%%%%%%%%%%
\section{Relating the propagator to fluid observables}
\label{sec:Observables}

Since our goal is to use the propagator formalism to obtain improved perturbative 
solutions for the fluid equations, we need a dictionary for \new{relating the propagator and wave function} 
to the classical fluid variables.
This is most conveniently done by relying on a phase-space formulation of quantum mechanics, which \new{translates propagator solutions for nonzero $\hbar$ to observables and is also ideally suited to determine the classical limits ($\hbar \to 0$).}

As we have introduced in Sec.\,\ref{sec:propagator}, the propagator evolves the wave function from its initial to the final state
\be
\label{eq:PsiKernel2}
  \psi(\vx;a) =  \int \! \dd^3 q \, K(\vx,\vq;a) \,  \psi^{\rm (ini)}(\vq) \,,
\ee 
with the initial condition $\psi(\vx;\text{\footnotesize $a\!=\!0$}) =\psi^{\rm (ini)}(\vq)$. We will consider wave functions constructed from propagators that satisfy the Schr\"odinger equation~\eqref{eq:Schroedi}. 
To establish a connection between the wave function~\eqref{eq:PsiKernel2} 
and the classical fluid variables,
we employ a method from quantum mechanics for studying quantum corrections to classical statistical mechanics --- which is a closely related problem. Following those ideas, we construct a phase-space distribution function $f(\vx,\vp)$ from the wave function $\psi(\vx)$ using the Wigner function \cite{Wigner32},\footnote{%
We note that the Wigner distribution function is technically
not a proper phase-space distribution when resolved on phase-space scales smaller than $\hbar$, since it can be negative and thus is a quasi-probability distribution that escapes the simple interpretation as a probability density. Hence, one should interpret equation~\eqref{eq:Wigner} in a coarse-grained sense avoiding violation of uncertainty relations, which can be formalised by using the Husimi distribution~\cite{Husimi40}. Since we will be interested in the classical limit, this coarse-graining scale will ultimately become superfluous.
}
which depends explicitly on a phase-space coarse-graining scale $\hbar$,
\be
\label{eq:Wigner}
f_{\rm W}(\vx,\vp) \! = \!\!\int \!\! \frac{\dd^3 x'}{(2\pi)^3}  \exp \!\!\left[ \!\dfrac{-\ii\vp \cdot \vx'}{a^{3/2}}\right] \!\psi(\vx+\tfrac{\hbar}{2} \vx') \,\bar\psi(\vx-\tfrac{\hbar}{2} \vx'),
\ee
where both $\psi$ and $f_{\rm W}$ are functions of time $a$, and $\bar\psi$ indicates the complex conjugated wave function. For convenience we have absorbed the particle mass $m$ in the parameter $\hbar$, and have included the factor $a^{-3/2}$ in front of the momentum $\vp$. This factor
stems from the fact that our wave function is defined in terms of a peculiar velocity that is related to the conjugate momentum via $a^{3/2}$ in EdS; see App.\,\ref{app:Schroedi} for details.

The way the Wigner distribution $f_{\rm W}(\vx,\vp)$ is constructed guarantees
that all phase-space information is encoded in the wave function. It is built in such a way, that the normalised density $\varrho \equiv 1+ \delta$
and the mean peculiar momentum $\v{j}=(1+\delta)\vv$
are obtained as the first two kinetic moments\footnote{Within the single particle probabilistic Copenhagen interpretation of quantum mechanics, $\rho$ is usually called the `probability density' and $\v{j}$ the (conserved) `probability flux'.}
\begin{align}
\label{eq:rhofromPsi}
  \varrho(\vx) &= \!\int \!\!\dd^3p\, f_{\rm W}(\vx,\vp) = |\psi|^2\,, \\
\label{eq:jfromPsi}
\v{j}(\vx) &= \!\int \!\!\dd^3p \frac{\vp}{a^{3/2}} f_{\rm W}(\vx,\vp) =\frac{\ii \hbar}{2}[\psi\vnabla \bar\psi-\bar\psi\vnabla\psi] \,.
\end{align} 
Note that the velocity can be written as a gradient of the phase, 
$\vv(\vx)=-\vnabla \phi_{\rm v}(\vx)$, of the wave function $\psi=\sqrt{1+\delta}\exp(-\ii\phi_{\rm v}/\hbar)$,
if and only if the amplitude and phase are sufficiently smooth. Shell-crossing however causes strongly oscillatory behaviour, see Fig.\,\ref{fig:free-sine-wave2}, which also generates vorticity as we shall discuss later in Secs.~\ref{sec:vorticity-theory} and~\ref{sec:vorticity-result}. This vorticity can be nonetheless extracted 
from the velocity field $\vv=\v{j}/\rho$ using the just introduced kinetic moments.

While in principle we could work with explicit expressions for density and momentum, the Wigner function provides a concise and elegant way of simultaneously encoding density and velocity information, which allows us to infer the Lagrangian displacement and corresponding velocity 
from its classical limit.

\subsection{Fluid variables from the classical limit}
Taking the classical limit
$\hbar \rightarrow 0$, after having obtained the solutions for the Wigner distribution~\eqref{eq:Wigner} for nonzero~$\hbar$, we obtain the phase-space distribution of a perfect fluid
\be
\label{eq:ffluid}
\lim_{\hbar \to 0} f_{\rm W}(\vx,\vp) = \varrho(\fett{x}) \, \delta_{\rm D}^{(3)}\left(\frac{\vp}{a^{3/2}}-\vv(\fett{x})\right) := f_{\rm fl}(\vx,\vp)\,,
\ee
with a velocity $\vv(\vx)$ that is single-valued before shell-crossing. \new{Note that the wave-function $\psi$ itself depends on $\hbar$, as illustrated by the split in amplitude and phase, $\psi=\sqrt{\varrho} \exp(-\ii\phi_{\rm v}/\hbar)$. Hence, the limit $\hbar\rightarrow 0$ needs to be taken with care and gives a nonzero peculiar velocity despite the $\hbar$ prefactor in Eq.\,\eqref{eq:jfromPsi}.} Using mass conservation~\eqref{eq:masscons}, we can 
formulate the distribution function of the perfect fluid in Lagrangian coordinates
\begin{align}
\label{eq:fLagrange}
\!\!\!f_{\rm fl}(\vx,\vp)=
\!\!\int\! \dd^3q\, \delta_{\rm D}^{(3)}\left[\vx-\vq-\v{\xi}(\vq)\right] \,\delta_{\rm D}^{(3)}\!\left[\frac{\vp}{a^{3/2}}-\vv^{\rm L}(\vq)\right] ,
\end{align}
where $\v{\xi}(\vq)$ is the Lagrangian displacement~\eqref{eq:defDisplacement} and $\vv^{\rm L}(\vq)=\vv(\vx(\vq;a);a)$ 
is the Lagrangian representation of the velocity evaluated at the Eulerian position $\fett{x}(\fett{q};a)$.
Hence, by performing the classical limit of the Wigner phase-space distribution~\eqref{eq:Wigner} for a given wave function, we can straightforwardly read off the corresponding Lagrangian displacement and velocity.

Let us demonstrate the outlined technique for obtaining the fluid variables, by using the free theory as an instructive example.
The corresponding wave function $\psi_0\!=\!\int \dd^3 q \, K_0(\vx,\vq,a) \,\psi^{\rm (ini)}(\vq)$, obtained from the free theory propagator $K_0$ from Eq.\,\eqref{eq:K0sol}, reads
\be 
\label{eq:psifree}
\!\!\!\!\!\psi_0(\vx; a)=\!\! \int\!\! \frac{\dd^3 q}{(2 \pi \ii \hbar a)^{\frac{3}{2}}} \exp \left[\frac{\ii(\vx-\vq)^2}{2\hbar a} - \frac{\ii}{\hbar} \varphi^{\rm (ini)}_{\rm g}(\vq) \right] \,,
\ee 
where the part
$\exp[ -\ii\varphi^{\rm (ini)}_{\rm g}(\vq)/\hbar] \equiv \psi^{\rm (ini)}(\fett{q})$
reflects the initial condition for the wave function,
 in accordance with the used boundary conditions~\eqref{slaving}.
Plugging $\psi_0$ into the Wigner distribution~\eqref{eq:Wigner}, we have three integrals over $\vx'$, $\vq$ and $\vq'$. 
The latter two integrals can be simplified with a change of variables, using
center of mass  $\fett{q}_+= (\vq+\vq')/2$ and difference coordinates $\fett{q}_-=\vq-\vq'$. We obtain 
\begin{align}
\label{eq:Wignerfree}
 \!\!\!\!f_{\rm W,0}&=\!\int\!\! \frac{\dd^3 x'}{(2\pi)^3}\!\! \int\!\frac{\dd^3q_+\, \dd^3 q_-}{(2\pi\hbar a)^3} \exp\!\left[\ii \vx' \cdot \left(\frac{-\vp}{a^{3/2}}+\frac{\vx- \fett{q}_+}{a}\!\right)\!\right] \nonumber \\
 &\hskip-0.2cm\times \exp\left\{\frac{-\ii}{\hbar a} \left[\fett{q}_- \cdot \left(\vx-\fett{q}_+\right) + a\,\delta\varphi(\fett{q}_+,\fett{q}_-)\right] \right\} \,,
\end{align}
where we have defined
\begin{align}
\label{eq:deltaphi}
\delta\varphi(\fett{q}_+,\fett{q}_-) &= \varphi^{\rm (ini)}_{\rm g}\left( \text{\footnotesize $\fett{q}_+ +$} \tfrac{\fett{q}_-}{2}\right)-\varphi^{\rm (ini)}_{\rm g}\left( \text{\footnotesize $\fett{q}_+ -$} \tfrac{\fett{q}_-}{2}\right) \,.
\intertext{Since we are considering the classical limit $\hbar \rightarrow 0$, the complex exponent in~\eqref{eq:Wignerfree} will vary very quickly for large $\fett{q}_-$ and cancel out their contribution.
Thus, in the classical limit the most dominant term in the integrand will come from terms for which $\fett{q}_-$ are small, thereby justifying to approximate
 $\delta \varphi$ in a leading-order Taylor expansion around small $\fett{q}_-$,}
   \delta\varphi &= \fett{q}_- \cdot \vnabla \varphi^{\rm (ini)}_{\rm g}(\fett{q}_+) + {\cal O}(q_-^3)\,.
\end{align}
In App.\,\ref{app:SPA} we show that this classical limit is closely related to the so-called stationary phase approximation.
Returning to the integrand, performing the integrations over $\vx'$ and $\fett{q}_-$, we then obtain 
\begin{align}
\label{eq:Wignerfreelimit}
 \lim_{\hbar \to 0} f_{\rm W,0} =  \int \dd^3 q\, \,\delta_{\rm D}^{(3)}\big[ \vx- \vq+a\vnabla\varphi^{\rm (ini)}_{\rm g}(\vq) \big]& \nonumber \\
   \times \,\delta_{\rm D}^{(3)} \left[ \frac{\vp}{a^{3/2}} +\vnabla \varphi^{\rm (ini)}_{\rm g}(\vq)\right]& \,,
\end{align}
where we have renamed the integration variable according to $\fett{q}_+ \to \fett{q}$ for convenience.
Comparing this to the 
fluid distribution function in Lagrangian coordinates, Eq.\,\eqref{eq:fLagrange}, we can read off the displacement and velocity
\begin{align}
\label{eq:displacevelLO}
\v{\xi}_{0}(\vq)=- a \vnabla \varphi_{\rm g}^{\rm (ini)}(\vq)\,,\quad 
\vv_{0} (\vq)= -\vnabla \varphi_{\rm g}^{\rm (ini)}(\vq) \,.
\end{align}
These solutions agree with those obtained from the ZA (cf.\ Eq.\,\eqref{eq:ZA}), 
and thus, in the classical limit \new{and to leading order in perturbation theory, 
we reproduce results from classical fluid dynamics. (See Eq.\,\eqref{NLOclassical} for  the classical limit at second order of our propagator method.)}

%we have established that our wave approach reproduces leading-order results from classical fluid dynamics.

Equipped with a method to relate semiclassical propagators and wave functions to fluid observables, we will proceed to perturbatively solve the Schr\"odinger equations for the propagator and translate our solutions to the Lagrangian displacement and velocity field in Sec.~\ref{sec:PPT}.

\subsection{The appearance of vorticity after shell-crossing}
\label{sec:vorticity-theory}

Under certain circumstances, the 
Kelvin-Helmholtz invariant $\Gamma$, given in Eq.\,\eqref{eq:Kelvin},  also persists for quantum and semiclassical systems.
In particular,
for sufficiently smooth initial conditions and by using a Madelung transformation, 
in Ref.\,\cite{Damski2003} it has been shown that $\Gamma$
is also an invariant under evolution with a quantum Hamiltonian (i.e.,  under the Schr\"odinger equation), if one simply replaces $\v{v}$ with~$\v{j}/\rho$ and ensures that the integral contour goes only through 
regions where the velocity is well defined in the course of the evolution.

In quantum systems, vorticity is quantised \cite{Gross1961,Feynman1958}. 
Since the wave function is always
single-valued, quantised vorticity can only arise from topological defects where the phase factor, $\phi_{\rm v}/\hbar$, undergoes a localised phase jump of integer multiples of $2\pi$. Since $\Gamma$ has to vanish for initially irrotational systems,
it is topologically required that vortices can only be produced in pairs \cite{Savchenko1999}, called {\it rotons}, i.e., 
 \be
\frac{1}{2\pi\hbar} \oint_{C(a)}\, \vnabla \phi_{\rm v}\cdot \dd\v{x} = n_+ - n_- = 0\,\,, \,\quad  n_\pm\in \mathbb N\,,
 \ee
and thus, the sum of negative $n_-$ and positive $n_+$ topological charges is conserved. 

Later in Sec.\,\ref{sec:Examples}
we show that at times shortly after shell-crossing,  where vorticity is generated,
we indeed observe the appearance of such rotons.

%%%%%%%%%%%%%%%%%%%%%%%%%%%%%%%%%%%%%%%%%%%%%%%%%%%%%%%%
\section{Perturbative treatment of the propagator}
\label{sec:PPT}

In the following, we will use SPT results for $V_{\rm eff}$, as discussed in Sec.\,\ref{sec:fluid} as an input to the propagator equation~\eqref{eq:Schroedi} and solve it in a perturbative fashion. From this perturbative solution, we will extract the Lagrangian displacement and velocity  
using the method described in Sec.~\ref{sec:Observables}.

Since we already know the solution in the absence of the effective potential, we split the nonlinear propagator into the free propagator and an exponential term 
\be
\label{eq:Kansatz}
   K(\vq,\vx; a) 
   =  K_{0}(\vq,\vx; a) \,
     \exp\left(  \frac \ii \hbar S_{\rm tid}(\vq,\vx; a)  \right) \,.
\ee
When combining the exponentials, one recognises the total action $S=S_0+S_{\rm tid}$ as a sum of the free particle contribution $S_0$ from Eq.\,\eqref{eq:S0} and the \new{tidal interaction terms encoded in $S_{\rm tid}$.} 
Plugging the {\it Ansatz}~\eqref{eq:Kansatz} for the total propagator $K$ into the evolution Eq.\,\eqref{eq:Schroedi}, and using Eq.\,\eqref{eq:freeSchroedi},
one obtains a differential equation for the interaction term $S_{\rm tid}= S_{\rm tid} (\vq, \vx; a)$
\begin{align}
\label{eq:evoE}
\!\!\hat D_{a}(\vq,\vx)\, S_{\rm tid}  -  \frac{\ii \hbar}{2} \vnablax^2 S_{\rm tid} +  \frac{(\vnablax S_{\rm tid})^2}{2} = - V_{\rm eff} \,, 
\end{align}
where $\hat D_{a}(\vq,\vx) \equiv \partial_a +  (1/a)[\vx-\vq] \cdot \vnablax$. The source term is given in terms of
a time-Taylor series for the effective potential
$V_{\rm eff}(\vx;a) \equiv \sum_{n=2}^\infty V_{\rm eff}^{(n)}(\vx) \,a^{n-2}$, and for the interaction part of the action one can impose a PT {\it Ansatz} 
\be \label{eq:intAns}
  S_{\rm tid}(\vq,\vx;a) = \sum_{n=1}^\infty S_n(\vq,\vx;a) \,,
\ee
where $S_n$ is ideally ${\cal O}(a^n)$. In the following section, we explicitly solve for the NLO part $S_1$, which is the leading-order contribution to $S_{\rm tid}$.

\subsection{Next-to-leading order propagator}
At next-to-leading order (NLO), the effective potential $V_{\rm eff}^{(2)}(\vx)$ is time-independent and given by Eq.\,\eqref{eq:Veff}. Hence, the NLO contribution to
$S_{\rm tid}$, called $S_1$, is expected to be of order $a$.
Since the derivative operator, $\hat D_a$, decreases the power of $a$ by one, the other two terms on the LHS of
Eq.\,\eqref{eq:evoE} are of higher order and do not contribute to $S_1$. The evolution equation~\eqref{eq:evoE} thus simplifies to
\be
 \hat D_a (\vq,\vx)\,S_1(\vq,\vx;a)  = - V_{\rm eff}^{(2)}(\vx) \,.
\ee
The solution of this equation is given by 
\begin{align}
\label{eq:S1sol}
S_1(\vq,\vx; a)&= -\int_0^a \dd a'\, V_{\rm eff}^{(2)}\left(\vq+\tfrac{a'}{a} [\vx-\vq]\right)\\
\notag &= - a \int_0^1 \dd\tau\, V_{\rm eff}^{(2)}\left(\vq+\tau [\vx-\vq]\right)\,,
\end{align}
which is an integral over the time-independent effective potential along a straight line connecting the initial position $\vq$ and final position $\vx$. Since it is impracticable to evaluate this integral at every point, let us make further approximations. Preferably, we want to preserve the symmetry between initial and final positions $\vx\leftrightarrow\vq$ when exchanging $a\leftrightarrow -a$, which implies a time-reversible propagation. To this end, we use a two-endpoint approximation 
\begin{align}
\label{eq:S1solVTV}
S_1(\vq,\vx; a) &= -\frac{a}{2} \left[V_{\rm eff}^{(2)}(\vq) +V_{\rm eff}^{(2)}(\vx)\right] \,,
\end{align} 
in accordance with the employed PT {\it Ansatz}~\eqref{eq:intAns}.
This approximation corresponds to a (numerical) kick-drift-kick scheme \new{with ${\cal O}(a^3)$ accuracy}, in which the potential $V_{\rm eff}^{(2)}$ is evaluated at the initial position $\vq$, the particle propagated to its final position $\vx$ and then the potential evaluated there.%
\footnote{Instead of employing an approximation that amounts to a kick-drift-kick scheme, we could have also used a midpoint approximation that would correspond to a drift-kick-drift scheme, where the particle is first propagated from $\vq$ to the midpoint $(\vq+\vx)/2$, the potential evaluated there, and then the particle is propagated from the midpoint to the endpoint $\vx$. In the present case, we prefer the kick-drift-kick procedure, because it updates the effective potential only once.}
Now, we can combine the solution $S_1$ from Eq.\,\eqref{eq:S1solVTV} with the free kernel $K_0$ from Eq.\,\eqref{eq:K0sol} into the NLO kernel $K_{\rm NLO}$ from Eq.\,\eqref{eq:Kansatz} and hence the NLO wave function using Eq.\,\eqref{eq:PsiKernel2} to get
\begin{subequations} \label{eqs:NLOresult}
\be
\label{eq:PsiNLOa}
  \psi_1(\vx; a) =  \int\! \dd^3 q \, K_{\rm NLO}(\fett{x},\fett{q};a) \,\psi^{\rm (ini)}(\fett{q}) \,,
\ee 
with the NLO propagator 
\be
\label{eq:PsiNLOb}
  K_{\rm NLO}(\fett{x},\fett{q};a) =  (2 \pi \ii \hbar a)^{-3/2}  \exp\left[ \frac{\ii}{\hbar} g(\vx,\vq;a) \right] 
\ee
and 
\be
  g(\vx,\vq;a) = \frac{(\vx-\vq)^2}{2a} -\frac{a}{2} \left[ V_{\rm eff}^{(2)}(\vq) + V_{\rm eff}^{(2)}(\vx) \right] \,,
\ee
\end{subequations}
where $V_{\rm eff}^{(2)}$ is given by Eq.\,\eqref{eq:Veff}.
The propagator has the structure of a numerical kick-drift-kick scheme,  
which will simplify the numerical implementation later on.
Equations~\eqref{eqs:NLOresult} constitute one of the main technical results of this paper. 
We note that
alternatively to the above derivation, one could also solve the Schr\"odinger equation~\eqref{eq:Schroedi} in operator notation, as shown in App.\,\ref{app:SchroediOperatorPT}.

\subsection{Next-to-leading order observables in the classical limit}
To extract the Lagrangian displacement and velocity from the NLO wave function, we proceed along the lines as described in Sec.\,\ref{sec:Observables}. In analogy to the free wave function~\eqref{eq:Wignerfree}, we get for $f_{\rm W,1} = f_{\rm W,1}(\fett{x},\fett{p};a)$
\begin{align}
\label{eq:WignerNLO}
 \!\!\!\!f_{\rm W,1}\! &=  \!\int\!\! \frac{\dd^3x'}{(2\pi)^3}\!\! \int\!\frac{\dd^3 q_+\, \dd^3 q_-}{(2\pi\hbar a)^3} \exp\left[\ii \fett{x}' \cdot \!\left(\frac{-\vp}{a^{3/2}}+\frac{\vx-\fett{q}_+}{a}\right)\right]\!\!\!\!  \nonumber  \\
\notag &\quad \times\exp\left\{\frac{-\ii}{\hbar a} \left[\fett{q_-}\cdot \left(\vx-\fett{q}_+\right) + a\,\delta\varphi(\fett{q}_+,\fett{q}_-)\right] \right\}\\
&\quad \times \exp\left\{\frac{-\ii a}{2\hbar} \left[\delta V_2(\fett{q}_+, \fett{q}_-) + \delta V_2(\vx,\hbar \fett{x}')\right]\right\} \,,
\end{align}
where, in addition to $\delta\varphi$ from Eq.\,\eqref{eq:deltaphi}, we have defined
\be
\label{eq:deltaV2}
\delta V_2(\vq_+,\v{q}_-)= V_{\rm eff}^{(2)}( \text{\footnotesize $\vq_+ +$} \tfrac{\v{q}_-}{2}) - V_{\rm eff}^{(2)}(\text{\footnotesize $\vq_+-$}\tfrac{\v{q}_-}{2}) \,.  
\ee
As argued before, in the classical limit, the dominant contributions to the integral come from the terms~\eqref{eq:deltaV2} for which $\fett{q}_-$ is small. Thus we can write
\be
  \delta V_2(\vq_+,\v{q}_-) = \v{q}_-\cdot\fett{\nabla} V_{\rm eff}^{(2)}(\vq_+) + {\cal O}(q_-^3) \,.
\ee
Using this, the integrations over $\vx'$ and $\v{q}_-$ are easily performed. 
After some straightforward calculations, we obtain
\begin{align}
 \lim_{\hbar \to 0} f_{\rm W,1} =  \int \dd^3 q\, \,\delta_{\rm D}^{(3)}\big[ \vx- \vq - \fett{\xi}^{\rm NLO}(\fett{q};a) \big]& \nonumber \\
   \times \,\delta_{\rm D}^{(3)} \left[ \frac{\vp}{a^{3/2}}-\fett{v}^{\rm L, NLO}(\fett{q};a)\right]& \,,
\end{align}
with the NLO displacement and (Lagrangian) velocity  
\begin{subequations} \label{NLOclassical}
\begin{align}
\label{eq:displaceVTV}
 \!\xi_i^{\rm NLO}&=
- a  \varphi_{{\rm g},i}^{\rm (ini)}  - \frac{a^2}{2}  V_{{\rm eff},i}^{(2)}  \,,  \\
\label{eq:velVTV}
\!v_{i}^{\rm L, NLO} &= 
 -\varphi_{{\rm g},i}^{\rm (ini)} - a  V_{{\rm eff},i}^{(2)} 
 +\frac{a^2}{2}  \varphi_{{\rm g},m}^{\rm (ini)} \,V_{{\rm eff}, mi}^{(2)} \,.
\end{align}
\end{subequations}
To arrive at the final expression for the velocity, we expanded
$V_{{\rm eff},i}^{(2)}(\vx) \simeq V_{{\rm eff},i}^{(2)}(\vq) -a \varphi_{,l}(\vq)V_{{\rm eff},il}^{(2)}(\vq)$
to the leading order in the displacement, in accordance with the expansion scheme employed for the classical limit. 
Evidently, the NLO displacement~\eqref{eq:displaceVTV} agrees with its classical counterpart (Eq.\,\eqref{eq:2LPTclass}). The NLO velocity, by contrast, 
differs from the second-order velocity in LPT which is $v_{i}^{\rm L, LPT}\simeq -\varphi_{{\rm g},i}^{\rm (ini)} - a  V_{{\rm eff},i}^{(2)}$, since the NLO velocity includes an additional term that within the LPT expansion would be considered as higher order.

We now prove that this additional term in the NLO velocity is crucial for maintaining the Hamiltonian structure of the system, and thereby not to excite spurious vorticity. We do this by using the Cauchy invariants as a diagnostic tool which we have introduced in Eq.\,\eqref{eq:cauchyLPT}. For this
we need the Lagrangian map and its time derivative from our NLO formalism, which are easily obtained from the above results; they read respectively
\begin{subequations} 
\begin{align}
 \!\!x_{l,j}^{\rm NLO}  &=  
 \delta_{lj} - a \,\varphi_{{\rm g},lj}^{\rm (ini)} - \frac{a^2}{2}  V_{{\rm eff},lj}^{(2)}  \,, \\
\!\!\dot x_{l,k}^{\rm NLO} &= 
  -\varphi_{{\rm g},lk}^{\rm (ini)} - a\,V_{{\rm eff},lk}^{(2)} +\frac{a^2}{2} (\varphi_{{\rm g},m}^{\rm (ini)}\, V_{{\rm eff},lm}^{(2)})_{,k} \,.
\end{align}
\end{subequations} 
Plugging these expressions into the Cauchy invariants formula, we obtain
\be \label{eq:nofakevort}
{\cal C}_i^{\rm NLO} = \varepsilon_{ijk} \, x_{l,j}^{\rm NLO} \, \dot x_{l,k}^{\rm NLO} = 0 + \mathcal O(a^3)\,,
\ee
and hence there is no vorticity generated at order $a^2$, in contrast to the 2LPT case, see Eq.\,\eqref{eq:cauchy2LPT}. 
\new{For a numerical analysis related to vorticity, see Sec.\,\ref{sec:vorticity-result}.}

In summary, we have thus established a direct correspondence between our semiclassical propagator method and classical fluid mechanics. In particular, the NLO displacement field coincides exactly with its LPT counterpart up to second order, whereas the NLO velocity receives naturally an additional term that is missing in LPT at this fixed perturbation order. However, as we claim, this additional term is crucial to respect the underlying Hamiltonian structure; as outlined in Sec.\,\ref{sec:fakevorticity}, ignoring this term could lead to the artificial generation of vorticity.

%%%%%%%%%%%%%%%%%%%%%%%%%%%%%%%%%%%%%%%%%%%%%%%%%%%%%%%%
\section{Results beyond 1D collapse}
\label{sec:Examples}

\begin{figure}
\includegraphics[width=0.87\columnwidth]{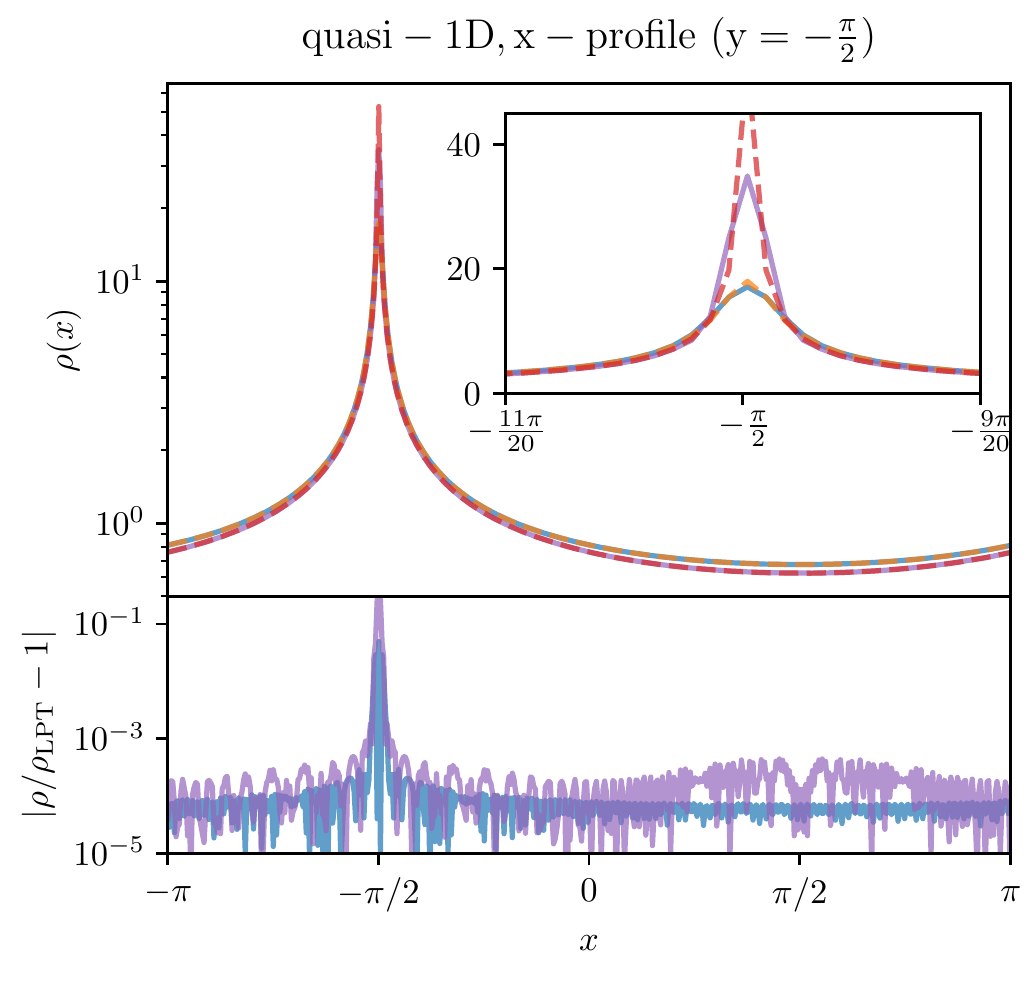}
\includegraphics[width=0.87\columnwidth]{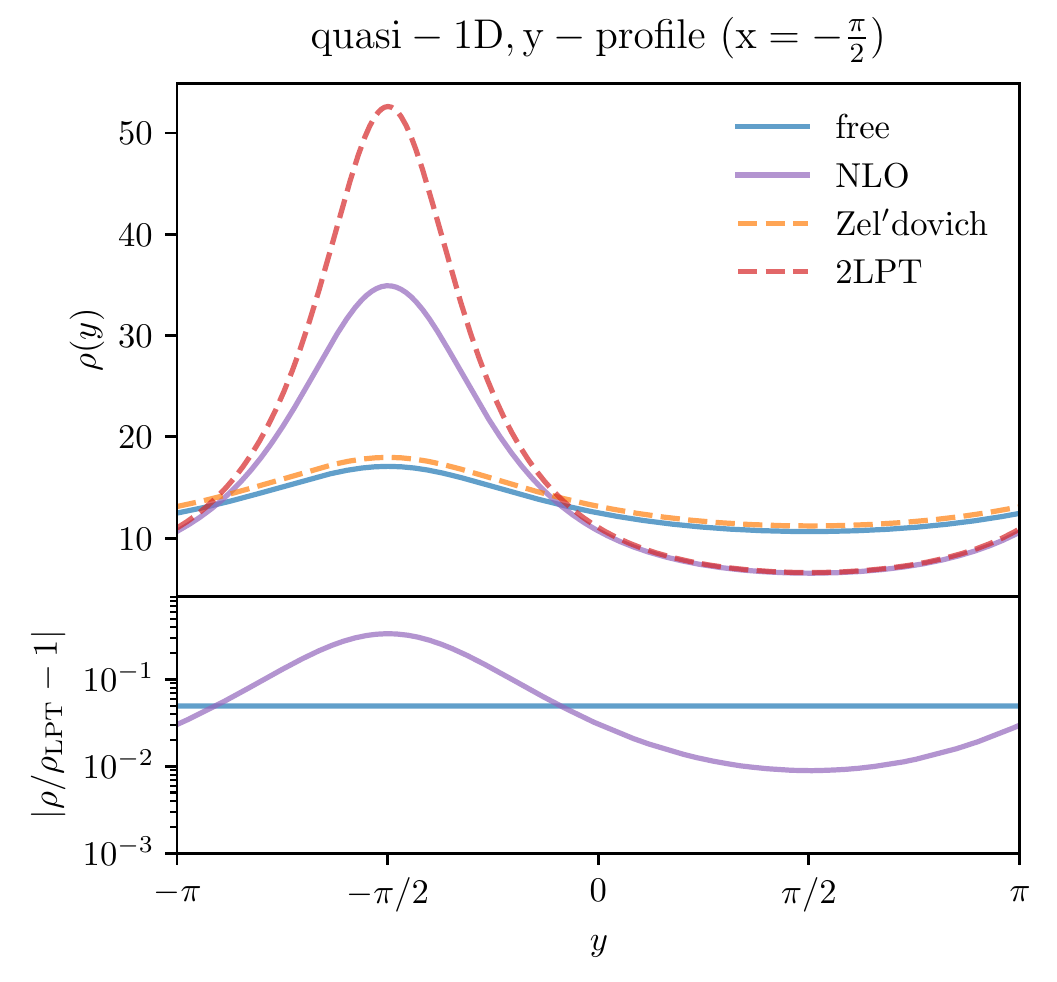}
\caption{Density profiles for quasi-one-dimensional collapse from the potential in Eq.~\eqref{eq:quasi1Dpot} 
%$\phi_{\rm v}^{\rm (ini)} = \sin q_1 + \epsilon \sin q_2$ with $\epsilon=1/4$
shortly before shell-crossing. The top figure shows the profiles in the 
orthogonal direction through the point of highest density on the ridge, while the lower figure shows the density profiles along the ridge where shell-crossing will happen shortly later.
We show the profiles for the classical results (dashed lines) and our propagator formalism (solid lines). In the lower panels we show the relative differences between the propagator and LPT densities (blue line: free/ZA; purple line: NLO/2LPT). The bottom figure demonstrates that the peak density along the (soon) shell-crossing ridge is regularised in the propagator formalism due to the finite $\hbar$. 
We have verified that mass conservation is satisfied to very high precision for all realisations.
} 
\label{fig:quasi_1d_profiles}
\end{figure}

\begin{figure*}
\includegraphics[width=\textwidth]{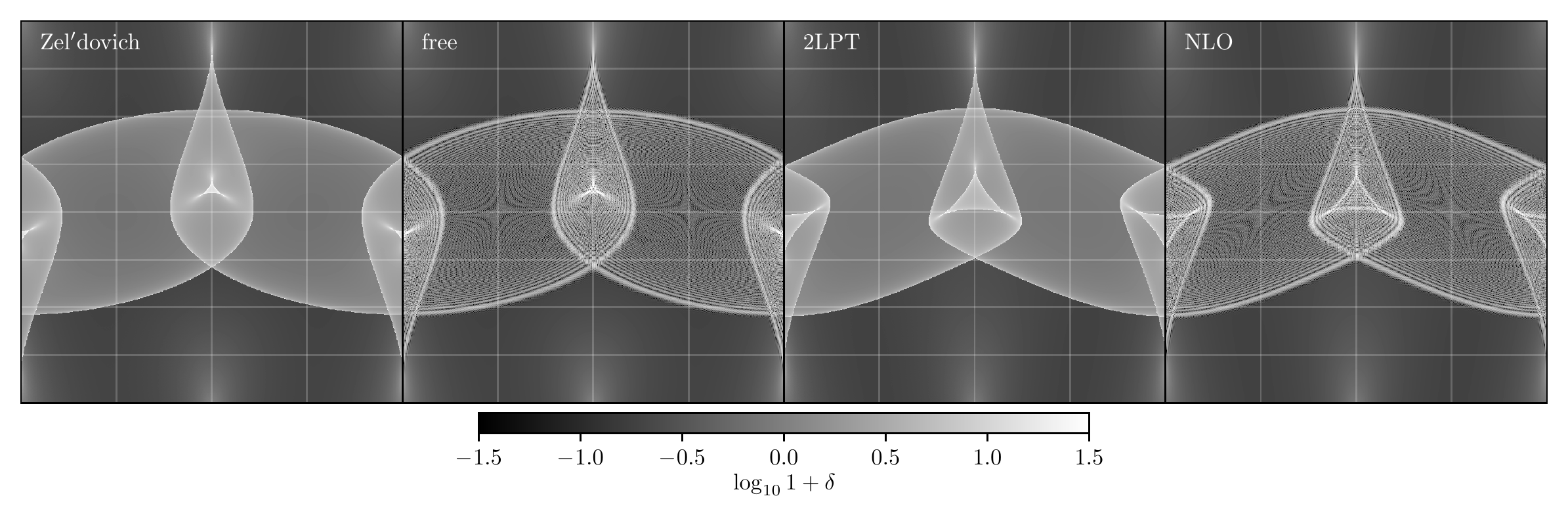}
\caption{Comparison of the density for the 2D phased wave problem with initial 
data given by Eq.\,\eqref{eq:phasedIC},
% $\phi_{\rm v}^{\rm (ini)} = -\cos\left( q_1 + \cos q_2\right)$ 
for Zel'dovich, the free propagator, 2LPT, and the NLO propagator (from left to right). Results are shown 
well after shell-crossing
%at $a=0.05$ 
for a box with $q_1,q_2\in[-\pi,\pi)$. We have used $1024^2$ particles to reconstruct the density on a grid of resolution $1024^2$ for the LPT figures, while the free and NLO propagators use $\hbar=5\times10^{-3}$ and are evaluated on a mesh of $1024^2$.} 
\label{fig:density-comparison}
\end{figure*}

\new{In the field of cosmological fluid dynamics, it is known that the effective gravitational potential, here dubbed $V_{\rm eff}$, is exactly zero if the initial conditions depend only on one space variable.
As a consequence, in 1D, the Zel'dovich solution  
becomes exact in the single-stream regime. Deviating from 1D, we have generically $V_{\rm eff} \neq 0$, and, as a result, the ZA performs rather poorly as gravitational tidal effects become non-negligible. In particular, this reflects in inaccurate predictions of the ZA for the shell-crossing time that worsen successively when deviating more and more from the 1D collapse \cite{Rampf17,Rampf:2017tne,Saga18}. Very similar performance issues are expected for the propagator method. Specifically, the free propagator~\eqref{eq:K0sol} is only accurate in the absence of tidal forces. 
Based on such considerations, we have determined the NLO propagator~\eqref{eq:PsiNLOb} in the presence of a nonzero effective potential by using perturbation theory.}

In the following, 
we  present numerical implementations for quasi-one-dimensional (Sec.\,\ref{sec:quasi1D}), as well as two-dimensional collapse problems (Sec.\,\ref{sec:2D}), both with an appropriate choice of tailored initial conditions (ICs). In both cases, 
we provide a quantitative comparison between the ZA and 2LPT predictions versus
 the results from the free and NLO propagators. 
 Numerical results for cosmological ICs will be investigated in a forthcoming paper.

Beyond 1D, the phenomenology of the flow is much richer and 
symmetry principles of the dynamics, such as the conservation of vorticity flux, become manifest.
Therefore, in Sec.~\ref{sec:vorticity-result}, we will analyse
 essential features of the vortical flow, which arise after shell-crossing and reflect 
the conservation of certain invariants.

%%%%%%%%%%%%%%%%%%%%%%%%%%%%%%%%%%%%%%%%%%%%%%%%%%%%%%%%%%%%%%%%%%%%%%%%%%%%%%%%%%%%%%%%%
%%%%%%%%%%%%%%%%%%%%%%%%%%%%%%%%%%%%%%%%%%%%%%%%%%%%%%%%%%%%%%%%%%%%%%%%%%%%%%%%%%%%%%%%%

\subsection{Quasi-one-dimensional collapse}
\label{sec:quasi1D}

For ICs that go beyond 1D,
higher-order effects for both the classical and propagator formalism become non-vanishing.
As outlined above, this is due to the fact that for departures from 1D the effective potential $V_{\rm eff}$ is generally nonzero. For  
an initial potential that is
(perturbatively) close to 1D, the bulk part stemming from $V_{\rm eff}$ is captured accurately by 2LPT and the NLO propagator.  
This is so, since the perturbation series converges very fast for such so-called quasi-1D ICs \cite{Rampf17,Saga18}. For quasi-1D ICs, we thus expect that the NLO propagator should deliver a very accurate description of the collapse.
To test the performance of the NLO propagator compared to 2LPT, we 
investigate quasi-1D ICs for the initial wave function 
$\psi^{\rm (ini)}(\vq)\!=\!\exp\left( - \ii \phi_{\rm v}^{\rm (ini)}(\vq)/\hbar\right)$, with the initial velocity potential
\begin{equation}
\label{eq:quasi1Dpot}
\phi_{\rm v}^{\rm (ini)}(q_1,q_2) = \sin q_1 + \epsilon \sin q_2 \,,
\end{equation}
where we have chosen $\epsilon =1/4$. 
We show the resulting density profiles shortly before shell-crossing in Fig.~\ref{fig:quasi_1d_profiles}. In the propagator case, 
we first evaluate the effective potential at initial position in real space,  propagate the particles in Fourier space, and then re-evaluate the potential at the final real-space position.
As mentioned above, this routine corresponds to a symplectic kick-drift-kick scheme  (see also App.\,\ref{app:SchroediOperatorPT}).
To obtain the realisation for the propagator, we have used a $512^2$ mesh to evaluate the solution with $\hbar=3\times10^{-2}$, while in the classical cases, we have used $256^2$ particles and evaluate the density field using the tessellation method of~\cite{Hahn2015} on a $512^2$ grid. We also show the relative differences between the 
predictions from LPT and the propagator method in the lower panels. 
Examining first the density profile along the main collapse direction $x$, one sees that all methods agree reasonably well with one another. This is expected since the perturbation along the perturbatively small
 second dimension is not large so that the evolution is still close to being one-dimensional,
for which the free and ZA solutions are good approximations.
The relative differences between free and ZA are at the accuracy level of our numerical experiments here at $\sim10^{-4}$, except at the location of future shell-crossing, where quantum corrections lead to a very localised large deviation at $\lesssim10^{-1}$ which is a direct result of the reduction of the peak density due to finite $\hbar$. 
The differences between NLO and 2LPT, by contrast, are much more prominent,  especially along the direction of
the density ridge (right panel in Fig.\,\ref{fig:quasi_1d_profiles}). 
For both classical and propagator methods, the peak densities at leading order and next-to-leading order are significantly different.
Apart from these differences at $x=-\pi/2$, the error in lower density regions is in fact mainly due to the linear interpolation used in the sheet reconstruction from the tessellation.

%%%%%%%%%%%%%%%%%%%%%%%%%%%%%%%%%%%%%%%%%%%%%%%%%%%%%%%%%%%%%%%%%%%%%%%%%%%%%%%%%%%%%%%%%
%%%%%%%%%%%%%%%%%%%%%%%%%%%%%%%%%%%%%%%%%%%%%%%%%%%%%%%%%%%%%%%%%%%%%%%%%%%%%%%%%%%%%%%%%

\begin{figure*}
\includegraphics[width=\textwidth]{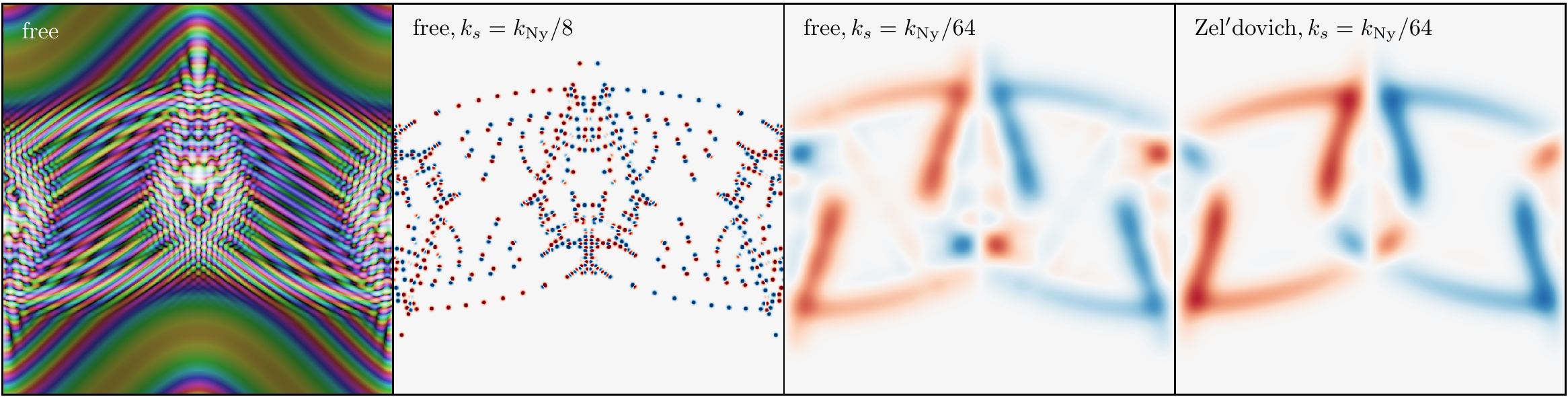}
\caption{The wave function $\psi$ (left panel, shown using domain coloring), as well as the vorticity $\fett{\omega}=\fett{\nabla} \times (\fett{j}/[1+\delta])$ (other panels) for the phased wave problem; initial data is provided by Eq.\,\eqref{eq:phasedIC}. The second and third panels from the left show the vorticity obtained using the free propagator, filtered with a Gaussian filter in Fourier space on scales of $1/8$ and $1/64$ the Nyquist wave number to highlight both the large-scale transversal modes and the topological defects from which they arise. The rightmost panel shows the corresponding vorticity using the Zel'dovich approximation with a smoothing to facilitate comparison to the large-scale free propagator case shown next to it. Time and initial conditions are identical to Fig.~\ref{fig:density-comparison}, but in order to highlight the role of $\hbar$, it has been increased to $\hbar=0.05$. The color scale for vorticity has been adjusted to highlight best the various features in each panel.} 
\label{fig:vorticity-comparison}
\end{figure*}

%%%%%%%%%%%%%%%%%%%%%%%%
\subsection{2D collapse -- the density field}\label{sec:2D}
\label{sec:phasedwave}

For reasons explained above, it is expected that the impact of $V_{\rm eff}$ becomes more prominent when departing greatly from 1D initial conditions. To demonstrate this we use in the present section the 2D initial potential
\begin{equation} \label{eq:phasedIC}
\phi_{\rm v}^{\rm (ini)}(q_1,q_2) = - 2 \cos\left( q_1 + \cos q_2\right),
\end{equation}
which represents a (strongly) phased plane wave (cf.\,\cite{Hahn2013}). In Fig.~\ref{fig:density-comparison}, 
we show the resulting two-dimensional density field for the four approaches 
\new{at $a=1$, i.e., well after shell-crossing.}
%at $a=0.05$, which is well after shell-crossing. 
For the Lagrangian perturbation theory results, we have used $1024^2$ 
fluid particles that have been evolved under the ZA or 2LPT in a single time step. In order to compute the multi-stream density field accurately, we have projected the tessellated dark matter sheet using the technique of~\cite{Abel2012} onto a uniform mesh of $1024^2$. 
For the propagator method, we have evolved the free and NLO
propagator directly on a $1024^2$ grid with 
\new{$\hbar\!=\!5 \times 10^{-3}$.}
%$\hbar\!=\!0.02$. 
Overall the results show excellent agreement between the ZA and the free propagator, as well as between 2LPT and NLO in terms of global shape of the caustics in the various multi-stream regions. Naturally, after shell-crossing, the propagator solutions show rapid oscillations that encode the multi-stream behaviour. This is related to the appearance of 
higher-dimensional caustics that translate into more complex diffraction patterns than
in the one-dimensional case (cf.\,\cite{Berry1980}). \new{When interpreted in a coarse-grained sense, the rapid oscillations disappear from the physical density and velocity, but encode the properties beyond the perfect pressureless fluid, in particular the vorticity that is induced by shell-crossing.}

\begin{figure*}
\includegraphics[width=\textwidth]{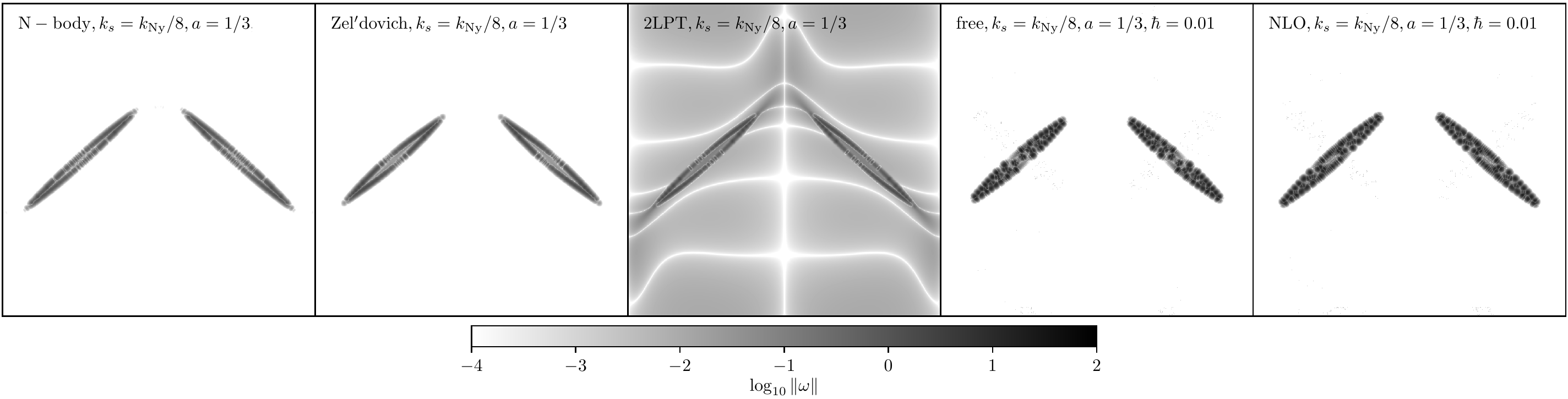}
\caption{\new{The norm of the vorticity $\fett{\omega}=\fett{\nabla} \times (\fett{j}/[1+\delta])$ for the phased wave problem (Eq.\,\eqref{eq:phasedIC}), shortly after the first shell-crossing (in the regions appearing as ``islands'' of vorticity generation). 
The panels display, from left to right, results from an $N$-body simulation, ZA, 2LPT, free theory and NLO. All results, except 2LPT, have in common a vanishing vorticity in single-stream regions (as it should), while 2LPT clearly exhibits spurious vorticity generation.} }
\label{fig:vorticity-spurious}
\end{figure*}

\subsection{2D collapse -- the vorticity}
% after shell-crossing}
\label{sec:vorticity-result}

As a final aspect of this paper, we  
investigate how vorticity arises 
in the classical and semiclassical picture, respectively.
For our propagator method, we determine the vorticity $\fett{w} \equiv \fett{\nabla} \times \fett{v}$ in Fourier space by first obtaining an expression for the velocity $\v{v} = \v{j}/(1+\delta)$, where the RHS is evaluated by using Eqs.\,\eqref{eq:rhofromPsi}--\eqref{eq:jfromPsi}. We thus calculate
\begin{equation}
\v{\omega} = \mathcal{F}^{-1}\left\{ -\ii \v{k} \times \mathcal{F}\left\{ \frac{\v{j}}{1+\delta} \right\}\right\},
\end{equation}
where $\mathcal{F}\{\cdot\}$ is a fast Fourier transform (FFT). Since the vortices are point-like, the inverse FFT produces heavy ringing, so that we have to additionally filter the vorticity fields. In order to avoid convolving transversal and longitudinal velocity components, we multiply with a Gaussian filter $\exp\left[ - k^2/k_s^2 \right]$, where $k_s$ is a filter scale, directly in Fourier space when also taking the cross product with~$\fett{k}$. 
For the LPT prediction of vorticity generation, by contrast, we use the method of \cite{Hahn2015} to explicitly carry out the multi-stream average.

%While we illustrate the results of this analysis for the ZA and the free propagator, the results for 2LPT and NLO are qualitatively identical. 

In Fig.~\ref{fig:vorticity-comparison}, we show the wave function using domain colouring in the left panel, along with the semiclassical vorticity for two different smoothing scales $k_s$ ($1/4$ and $1/16$ of the Nyquist wave number) in the two middle panels. When using the smaller smoothing scale (second panel from left), one can clearly see that the vortices are indeed point-like objects in two dimensions, which have a positive (red) or negative (blue) sign, and are concentrated around the caustics. In comparison with the full wave function (leftmost panel), one sees that the vortices are always associated with dark regions, where the amplitude of the wave function vanishes. As discussed in Sec.~\ref{sec:vorticity-theory}, vorticity is conserved and has thus to be pair-produced with opposite topological charge. These neighbouring positive and negative vortices are clearly visible in Fig.~\ref{fig:vorticity-comparison}. For the larger softening (third panel from left), it becomes obvious that, when averaging over multiple such quantum vortices, one obtains a 
large-scale limit which is very similar to the vorticity pattern obtained for the ZA (rightmost panel). The agreement of the properties of the two-dimensional flow between classical and quantum dynamics after filtering has been discussed in detail for the cosmological Schr\"odinger equation by~\cite{Kopp17}.

Remarkably, the propagator method allows us to explicitly predict the generation of vorticity without requiring a  
numerical solution of the Schr\"odinger equation. 
Furthermore, the propagator method does not  require multi-stream averaging.
This should be contrasted to the classical (multi)-fluid picture, where multi-stream averaging is mandatory \cite{Hahn2015} and computationally involved, even for simple cases like the ZA \cite{Pichon99vorticity}.

\new{Finally, in Fig.\,\ref{fig:vorticity-spurious}, we display the vorticity at times shortly after the first shell-crossing for the smoothing scale $k_s= k_{\rm Ny}/8$.
For reasons of comparison we have also performed an $N$-body simulation (leftmost panel)
which has been initialised at $a_{\rm ini}=1/30$. % using 2LPT initial conditions. 
In all panels the generation of vorticity through multi-streaming is visible by the appearance of ``vorticity islands'', while 2LPT predicts the spurious generation of vorticity in single-stream regions (see Sec.\,\ref{sec:fakevorticity} for a related discussion). We also note that the white lines in the 2LPT panel indicate where the spurious vorticity changes sign. Our NLO results, by contrast, are free of spurious vorticity; see e.g.\ Eq.\,\eqref{eq:nofakevort} for the provided theoretical argument.}

%%%%%%%%%%%%%%%%%%%%%%%%%%%%%%%%%%%%%%%%%%%%%%%%%%%%%%%%
\section{Conclusions}
\label{sec:conclusion}

{\it A. Summary.} We have introduced a novel semiclassical method for evolving CDM that, \new{for nonzero $\hbar$,} 
is free from any singular behaviour near the crossing of particle trajectories (shell-crossing). 
The key quantity of our method is the propagator, which fulfils a Schr\"odinger equation and encodes the transition amplitude of the wave function for the particle trajectories. This propagator can be determined by using perturbation theory. In a suitable coordinate system,  
the leading-order propagator dictates a ballistic motion of the particles with prescribed velocity,
which amounts to the classical Zel'dovich approximation.

To incorporate gravitational interactions in the propagator,
we have motivated the inclusion of an effective potential in the Schr\"odinger equation, which 
is also present in the cosmological fluid equations [Eq.\,\eqref{eq:Veffdef}].
Using standard perturbative techniques for the effective potential as an input for the Schr\"odinger equation, we have solved the associated propagator equation at next to leading order (NLO). The NLO solution for the propagator delivers the associated wave function, Eq.\,\eqref{eqs:NLOresult}, from which the NLO density and velocity are easily obtained [Eqs.\,\eqref{eq:rhofromPsi}--\eqref{eq:jfromPsi}].

By performing the classical limit, we have shown that our NLO result returns a displacement field~\eqref{eq:displaceVTV} and a corresponding density that are in agreement with the ones from Lagrangian perturbation theory (LPT) up to second order. The associated NLO velocity~\eqref{eq:velVTV} receives naturally an additional term ${\cal O}(a^2)$, that in LPT would be considered as third order, but  is actually important to preserve the underlying Hamiltonian structure of the system and to avoid a spurious excitation of vorticity in certain implementations of second-order LPT (see Sec.\,\ref{sec:fakevorticity}).

We compared our NLO results from the propagator method against LPT for  
two types of initial conditions, and find overall good quantitative agreement, see \mbox{Figs.\,\ref{fig:quasi_1d_profiles}--\ref{fig:density-comparison}.}
The propagator method regularises classical caustics, which leaves subtle imprints while preserving an overall good agreement in their global shapes and positions.
Furthermore, based on our analytical solutions for the propagator, we have demonstrated
that our method is 
\new{free of spurious vorticity generation (Fig.\,\ref{fig:vorticity-spurious}), as well as is} 
capable to predict the generation of vorticity after shell-crossing (Fig.\,\ref{fig:vorticity-comparison}). 
Although \new{the latter vorticity}
indeed arises solely through multi-stream dynamics, in our propagator method no explicit multi-stream averaging is required (see Sec.\,\ref{sec:vorticity-result} for a related discussion). In our formalism, vorticity manifests by the pair creation of topological defects, usually called rotons, along classical caustics.

{\it B. Outlook.} For simplicity, in the present work we have applied our propagator method to two-dimensional collapse problems only, however all provided tools are ready to use for full 3D calculations. Surely, the phenomenology for cosmological ICs will be much richer than explored for the present case studies, and therefore will be investigated in a forthcoming paper.
Another interesting avenue would be to include higher-order corrections in the propagator. For this one requires the external potential to third order, which we discuss in App.\,\ref{app:Veff}. 
Finally, our propagator method already shed some light on the highly complicated regime shortly after shell-crossing (including the generation of vorticity, velocity dispersion, etc.). For computations well beyond shell-crossing, however, the effective potential should be updated in order to grasp the full-fledged multi-stream regime, which is required to approach virialisation.

\section*{ACKNOWLEDGMENTS}
We thank Diego Blas, Lam Hui, Michael Kopp, \mbox{Giorgio} Krstulovic and Aseem Paranjape for many useful discussions and comments on the draft. We also thank the Wolfgang Pauli Institute, and in particular Norbert Mauser, for hospitality during the initial stages of this work and the participants of the workshop on `Numerics in cosmology: around the Schr\"odinger method' for interesting discussions. 
CU kindly acknowledges funding by the STFC grant no.\ RG84196 `Revealing the Structure of the Universe'. 
CR thanks the Lagrange Laboratory of the Observatoire de la C\^ote d'Azur for financial support.
CR acknowledges funding from the People Programme (Marie Curie Actions) of the European Union H2020 Programme under Grant Agreement No.\ 795707 (COSMO-BLOW-UP).
MG acknowledges support from the Marsden Fund of the Royal Society of New Zealand and from the European Research Council under the European Union's Seventh Framework Programme  (FP/2007-2013)/ERC Grant Agreement No. 616170.
OH acknowledges funding from the European Research Council (ERC) under the European Union's Horizon 2020 research and innovation programme (Grant Agreement No.\,679145, project 'COSMO-SIMS').

\appendix

%%%%%%%%%%%%%%%%%%%%%%%%%%%%%%%%%%%%%%%%%%%%%%%%%%%%%%%%
\section{Effective potential from standard perturbation theory}
\label{app:Veff}
Here we will show how the standard recursion relations from SPT can be used to determine the effective potential~\eqref{eq:Veffdef}, appearing in the Bernoulli equation~\eqref{eq:Bernoulli}, in an $a$-time expansion~\eqref{eq:Veffexpansion}. For notational simplicity, we use the following shorthand notations $\varphi_{\rm g}^{\rm (ini)} \rightarrow \varphi^{\rm (ini)}$ for the initial gravitational potential, and $\phi_{\rm v}\rightarrow \phi$ for the velocity potential.

The recursion relations for the perturbative expansion of the density contrast $\delta$ and velocity potential $\phi$ from Eq.\,\eqref{eq:Veffexpansion} give the following first two terms \cite{BernardeauReview} 
\begin{align}
  \!\!\phi^{(1)} &= \varphi^{\rm (ini)} \,, \label{phi1} \\
  \!\!  \delta^{(1)} &= \fett{\nabla}^2 \varphi^{\rm (ini)} \,, \label{delta1}\\
  \!\!\phi^{(2)} &= \,\nabla^{-2} \left[  \frac 3 7  \varphi_{,ll}^{\rm (ini)} \varphi_{,mm\phantom{l}}^{\rm (ini)}  + \varphi_{,llm}^{\rm (ini)} \varphi_{,m\phantom{l}}^{\rm (ini)} + \frac 4 7 \varphi_{,lm}^{\rm (ini)} \varphi_{,lm}^{\rm (ini)} \right] , \label{phi2}\\
  \!\!\delta^{(2)} &= \frac 5 7 \varphi_{,ll}^{\rm (ini)} \varphi_{,mm}^{\rm (ini)} +
    \varphi_{,llm}^{\rm (ini)} \varphi_{,m}^{\rm (ini)}
   + \frac 2 7   \varphi_{,lm}^{\rm (ini)} \varphi_{,lm}^{\rm (ini)} \,, \label{delta2}
\end{align}
where the derivatives and dependences are w.r.t.\ Eulerian coordinates.
To get an expression for $V_{\rm eff}^{(n)}$, we simply plug $\phi^{(n)}$ into the perturbed Bernoulli equation~\eqref{eq:Bernoulli}, which can be written as
\be
 V_{\rm eff}^{(n)}  = \partial_a \phi^{(n)} - \nabla^{-2} \sum_{s_1+s_2=n} \frac 1 2  \left( \phi_{,l}^{(s_1)} \phi_{,l}^{(s_2)} \right)_{,mm} \,.
\ee
 The first three solutions are 
\begin{align}
  V_{\rm eff}^{(1)} &= 0 \,, \label{eq:Veff1} \\
  V_{\rm eff}^{(2)} &= \frac 3 7 \nabla^{-2} \left[  \varphi_{,ll}^{\rm (ini)}  \varphi_{,mm}^{\rm (ini)} -  \varphi_{,lm}^{\rm (ini)}  \varphi_{,lm}^{\rm (ini)}\right]    \label{eq:Veff2} \,,\\
\notag V_{\rm eff}^{(3)} &=  \frac 1 3  \nabla^{-2}\Big[  \varphi_{,ll}^{\rm (ini)} \left(  \nabla^{-2} \delta^{(2)} +  \phi^{(2)} \right)_{,mm} 
   \\
\notag &\qquad\qquad -  \varphi_{,lm}^{\rm (ini)}  \left( \nabla^{-2} \delta^{(2)} + \phi^{(2)} \right)_{,lm} \Big]\\
   &+ \frac 1 9 \nabla^{-2}\left[  \varphi_{,llm}^{\rm (ini)} V_{{\rm eff},m}^{(2)} -  \varphi_{,m}^{\rm (ini)}   V_{{\rm eff},llm}^{(2)} \right]\,. \label{eq:Veff3}
\end{align}
Note that $V_{\rm eff}$ in the expansion~\eqref{eq:Veffexpansion} is only time-independent up to second order, while the third-order term is proportional to $a$-time $aV_{\rm eff}^{(3)}$.

As mentioned earlier, before shell-crossing, one can transform the Schr\"odinger equation~\eqref{eq:Schroedi} into a fluid-like system using a polar form for the wave function $\psi=\sqrt{1+\delta}\exp(-\ii\phi/\hbar)$ \cite{Madelung27}. One obtains a system equivalent to \eqref{eqs:fluid}, but with an extra term in the Bernoulli equation~\eqref{eq:Bernoulli} that adds to the effective potential
\be
\label{eq:QP}
V_{\rm eff}\rightarrow  V_{{\rm eff},\psi} = V_{\rm eff} + \frac{\hbar^2}{2} \frac{\fett{\nabla}^2\sqrt{1+\delta}}{\sqrt{1+\delta}}\,.
\ee
Note that, if one perturbatively solves the corresponding wave-mechanical fluid equations, in analogy to SPT for the fluid equations, one obtains identical solutions up to second order. The reason is that the leading order correction term in \eqref{eq:QP} is proportional to $\fett{\nabla}^2 \delta \propto a$ and hence only enters in the effective potential at third order, where it modifies~\eqref{eq:Veff3} to
\be
\label{eq:Veff3QP}
V_{{\rm eff},\psi}^{(3)} = V_{\rm eff}^{(3)} + \frac{\hbar^2}{12} (\fett{\nabla}^2)^2 \varphi^{\rm (ini)}\,.
\ee

%%%%%%%%%%%%%%%%%%%%%%%%%%%%%%%%%%%%%%%%%%%%%%%%%%%%%%%%

\section{Relation between the propagator formalism and Schr\"odinger-Poisson}
\label{app:Schroedi}
While the Schr\"odinger equation~\eqref{eq:Schroedi} used here might look similar to the Schr\"odinger method \cite{Widrow1993,Uhlemann2014} for approximating classical dynamics through the quantum-classical correspondence, they are physically distinct and rely on different assumptions, as we show now. The cosmological Schr\"odinger-Poisson equation in the formulation of Widrow \& Kaiser \cite{Widrow1993} reads
\be \label{eq:SPcosmic}
  \ii \hbar \partial_t \tilde \psi = - \frac{\hbar^2}{2a^2} \fett{\nabla}_x^2 \tilde \psi +  V \tilde \psi \,, \quad \fett{\nabla}_x^2 V =  \frac 3 2\frac{|\tilde \psi|^2-1}{a} \,,
\ee
where for simplicity we have absorbed the mass in $\hbar$ and set $4\pi G\bar \rho_0=3/2$. To distinguish the wave function in the different formulations, we attach a tilde to the wave function in the Widrow \& Kaiser formulation. The Schr\"odinger-Poisson equation can be regarded as an approximate treatment of the phase-space dynamics described by the Vlasov-Poisson equation \cite{Uhlemann2014,Uhlemann18}. When solved numerically, it provides a field-based method that complements particle-based $N$-body simulations and whose accuracy is controlled by the phase-space resolution $\hbar$ \cite{Kopp17,Garny18,Mocz18}. Additionally, the Schr\"odinger-Poisson equation is a physical model for scalar field (or wavelike) dark matter such as ultralight axions, where it has attracted considerable attention \cite{Schive14,Niemeyer16,Hui17,Hui18,Niemeyer18,Edwards:2018ccc} recently.

Evidently, the physical time Schr\"odinger equation~\eqref{eq:SPcosmic} does contain a gravitational potential $V$, but not a difference of a gravitational and velocity potential as our effective potential $V_{\rm eff}$. This is related to the fact that the associated wave function encodes the conjugate momentum. Employing the Madelung split, we can write the wave function as $\tilde \psi = \sqrt{1+\delta} \exp(-\ii \tilde \phi_{\rm p}/\hbar)$, where the phase $\tilde\phi_{\rm p}$ is the potential of the conjugate momentum. Since the conjugate momentum is defined as $\vp/m=a^2\dd\vx/\dd t$, which is related to our peculiar velocity $\vv=\dd\vx/\dd a$ via $\vp/m= \vv a^2 \dd a/\dd t =a^{3/2}\vv$ in an EdS universe, the conjugate momentum potential is related to our peculiar velocity potential as $\tilde \phi_{\rm p} = a^{3/2} \phi_{\rm v}$.  Substituting this relation in the wave function one has $\tilde \psi = \sqrt{1+\delta} \exp(-\ii a^{3/2} \phi_{\rm v}/\hbar)$. 

Similarly as done before, we can convert the Schr\"odinger equation for $\tilde \psi$ into fluid-type equations. Separating real and imaginary parts of Eq.\,\eqref{eq:SPcosmic}, leads to the fluid equations in physical time (see Eqs.\,(14)  in~\cite{Uhlemann2014}). Those fluid equations can be rewritten in $a$-time to obtain a fluid-like system~\eqref{eqs:fluid} with a modified effective potential that sources the Bernoulli equation~\eqref{eq:Bernoulli} for $\phi_{\rm v}=a^{-3/2}\tilde \phi_{\rm p}$ according to
\be
\label{eq:QP-WK}
V_{\rm eff}\rightarrow V_{{\rm eff},\tilde\psi} = V_{\rm eff} + \frac{\hbar^2}{2a^3} \frac{\nabla^2\sqrt{1+\delta}}{\sqrt{1+\delta}}\,,
\ee
which carries an extra $a^{-3}$ dependence in the $\hbar^2$-dependent `quantum' term in the effective potential, as compared to our expression~\eqref{eq:QP}. Using the peculiar velocity potential, we obtain an effective potential in an $a$-time Schr\"odinger equation for $\psi=\sqrt{1+\delta}\exp(-\ii\phi_{\rm v}/\hbar)$, but since the quantum potential term has a different time-dependence, it cannot be absorbed in the Laplacian, as was the case for our Schr\"odinger equation~\eqref{eq:Schroedi}. One can view the difference in the fluid equations, or the effective potential, as a time-dependent phase-space coarse-graining scale $\hbar$ (or a time-dependent mass). This property of a time-dependent coarse-graining  persists in the full-fledged multi-stream regime, where the evolution is governed by the Vlasov-Poisson equations, see \cite{Uhlemann18}.

\section{Stationary-phase approximation}\label{app:SPA} 

The stationary phase approximation (SPA) \cite{HoermanderBook} can be used to estimate integrals of the following type
\be
\label{eq:SPAint}
I(\lambda)=\left(2\pi \lambda\right)^{-\frac{3}{2}}\int \dd^3q\, h(\vq) \exp[\tfrac{\ii}{\lambda} g(\vq)]\,.
\ee
In the present context of the paper, evaluating such integrals in the vicinity of $\lambda \to 0$ is relevant when performing the classical limit.

For $\lambda \to 0$, the SPA states that the dominant contribution from the above integral comes from
\be 
\label{eq:SPA}
\lim_{\lambda\rightarrow 0} I(\lambda) = \sum_{\vq_c} \frac{h(\vq_c)\exp[ \tfrac{\ii}{\lambda} g(\vq_c)] }{\left|\det\left(H_{ij}\right)\right|^{1/2}}  \exp\left(\frac{\ii\pi}{4}\text{sign}\left(H_{ij}\right)\right)\,,
\ee
where one sums over all critical points $\vq_c$ for which  the first-order Taylor coefficient $\partial g(\fett{q})/\partial q_i |_{\vq= \vq_c}$ vanishes and $H_{ij}= \partial^2 g(\fett{q})/(\partial q_i\partial q_j) |_{\vq= \vq_c}$ is the Hessian. The signature of the Hessian, which is the difference between the number of negative and positive eigenvalues, determines the prefactor in the last exponential of~\eqref{eq:SPA}. Note that this is just a Wick rotated version of the formula for the method of steepest descent, or so-called saddle-point approximation.

\subsection*{Application to free theory observables}

Let us apply the SPA to the free wave function~\eqref{eq:psifree}.
This wave function has the form~\eqref{eq:SPAint} with the constants $\lambda=a\hbar$, $h_0^{\psi}=\exp(-3\ii\pi/4)$ and the exponential
\be
g_0(\vx,\vq) = \frac{(\vx-\vq)^2}{2} - a\varphi(\vq)\,.
\ee
Considering the classical limit $\hbar\rightarrow 0$, we obtain the condition for a point of stationary phase
\be \label{eq:SPAexp}
 \vnabla_{\vq} g_0(\vx,\vq_c) \stackrel ! = 0 \ \Rightarrow \ \vq_c  = \fett{x} + a \nabla \varphi(\vq_c)    \,,
\ee 
which implicitly determines the critical point $\vq_c$. Before shell-crossing, for every $\vx$ there exists only one such critical point $\vq_c$.
To perform the integral over $\vq$ in~\eqref{eq:psifree}, we expand $g_0(\vx,\vq)$ in a Taylor series up to quadratic order around the critical point $\vq_c$ 
\begin{align}
   g_0(\vq) &\simeq \frac 1 2  (a \nabla \varphi(\vq_c))^2 - a \varphi(\vq_c) \nonumber \\
    &\quad + \frac 1 2 \left[ \delta_{ij} - a \varphi_{,ij}(\vq_c)\right] (q_i -q_{c,i}) (q_j - q_{c,j}) \,.
\end{align}
Then, one can shift the integration $\vq\rightarrow \tilde{\vq} = \vq - \vq_c$ and perform the following Gaussian integral
\begin{align}
\notag \int \frac{\dd^3 \tilde q}{(2 \pi \ii \hbar a)^{\frac{3}{2}}} \, \exp\left\{ \frac{\ii}{2\hbar a} \left[ \delta_{ij} - a \varphi_{,ij}(\vq_c) \right] \tilde q_i \tilde q_j \right\} \\
=\frac{1}{\sqrt{\det[\delta_{ij}-a\varphi_{,ij}(\vq_c)]}}\,.
\end{align}
The corresponding wave function  is 
\be
\label{eq:Psi0SPA}
\psi_0^{\rm SPA}(\vx) =  \frac{\exp\left[ \frac{\ii}{\hbar} \left( \tfrac{1}{2}(a\nabla\varphi(\vq_c))^2 -a\varphi(\vq_c)\right)\right]}{\sqrt{\det[\delta_{ij}-a\varphi_{,ij}(\vq_c)]}}\,,
\ee
where $\vq_c=\vq_c(\vx)$ according to the stationary phase condition~\eqref{eq:SPA}. From this, we can easily compute the density according to the definition~\eqref{eq:rhofromPsi}, which reads
\begin{align}
\label{eq:deltaSPA}
\delta_0(\vx) +1 &=|\psi_0(\vx)|^2= \frac{1}{\det[\delta_{ij}-a\varphi_{,ij}(\vq_c)]}  \nonumber \\
& =\int \dd^3q\, \delta_{\rm D}^{(3)} \left[ \vx-\vq+a\vnabla\varphi(\vq) \right] \,,
\end{align}
from which one evidently recovers the Zel'dovich displacement field~\eqref{eq:displacevelLO}. 
By exactly the same arguments, the SPA for the NLO wave function delivers the 2LPT displacement field.

To derive the velocity, one needs a SPA for the spatial gradient of the wave function. In this case one has the prefactor $h_0^{\nabla \psi}= \ii(\vx-\vq)\exp(-3\ii\pi/4)/(a\hbar)$, which is evaluated using the  stationary phase condition~\eqref{eq:SPA} to obtain
\be
\label{eq:nablaPsi0SPA}
\vnabla\psi_0^{\rm SPA}(\vx) = \frac{-\ii\vnabla\varphi(\vq_c) }{\hbar} \frac{\exp\left( \frac{\ii}{\hbar} \left[ \tfrac{[a\nabla\varphi(\vq_c)]^2}{2} -a\varphi(\vq_c)\right]\right)}{\sqrt{\det[\delta_{ij}-a\varphi_{,ij}(\vq_c)]}}\,.
\ee
From this expression, we can now compute the velocity according to Eq.\,\eqref{eq:jfromPsi} to get
\begin{align}
\label{eq:vSPA}
\vv_0(\vx)&=\frac{\ii\hbar}{2}\frac{[\psi_0\vnabla\bar\psi_0-\bar\psi_0\vnabla \psi_0](\vx)}{1+\delta_0(\vx)}=-\vnabla\varphi(\vq_c)\,,
\end{align}
which agrees with the previous result from Eq.\,\eqref{eq:displacevelLO} for the velocity in Lagrangian coordinates.\\[0.4cm]

\section{Perturbative propagator in operator notation}
\label{app:SchroediOperatorPT}
A particularly concise perturbative expansion can be obtained from an operator expansion. To solve the Schr\"odinger equation~\eqref{eq:Schroedi} in operator notation, we write the Hamiltonian operator~\eqref{eq:Hop} at leading order, where it is time-independent
\be
\hat H^{(2)} = -\frac{\hbar^2}{2}\fett{\nabla}_x^2 + V_{\rm eff}^{(2)}(\vx) =: \hat T + \hat V\,.
\ee
For simplicity, we denote the operators associated to kinetic and potential energy as $\hat T$ and $\hat V$, respectively. Since neither of those operators have explicit time dependence, the Schr\"odinger equation~\eqref{eq:Schroedi} can be integrated
\begin{equation}
\psi(\vx,a) = \exp\left[ -\frac{\ii}{\hbar} a\left(\hat{T}+\hat{V}\right)\right]\,\psi^{\rm (ini)}(\vx)\,.
\end{equation}
The Baker-Campbell-Hausdorff formula (or the equivalent reverse Zassenhaus formula) allows to express exponentials of sums of operators as products of exponentials according to
\be
\label{eq:BCH}
\e^{\epsilon(\hat T+\hat V)}=\e^{\epsilon\hat T} \e^{\epsilon\hat V} \exp\left(-\tfrac{\epsilon^2}{2} [\hat T,\hat V]+\mathcal O(\epsilon^3)\right)\,,
\ee
where $\epsilon=-\ii a/\hbar$, 
and higher-order terms come from nested commutators which are denoted with $[\cdot,\cdot]$. Using a threefold decomposition of the Hamiltonian $\hat H$ in $VTV:= \hat V/2 + \hat T + \hat V/2$ allows to both arrive at a time-symmetric formula {\em and} cancel all even-order correction terms (and thus the leading order is proportional to $\epsilon^2 \sim a^2$)
\be
\label{eq:BCHVTV}
\exp\left(\epsilon(\hat T+\hat V) +\mathcal O(\epsilon^3)\right)= \e^{\epsilon \hat V/2} \, \e^{\epsilon\hat T} \,\e^{\epsilon\hat V/2}\,.
\ee
This provides another motivation for the VTV approximation of~\eqref{eq:S1solVTV}, which leads to the NLO propagator~\eqref{eq:PsiNLOb}. The approach outlined above is of course equivalent also to the usual operator-split approach followed by symplectic schemes when numerically integrating classical Hamiltonian systems, cf.\ \cite{Yoshida:1990}.

%\newpage

\bibliography{database}

\end{document}